\documentclass[graybox]{svmult}

\usepackage{mathptmx}       
\usepackage{helvet}         
\usepackage{courier}        
\usepackage{type1cm}        
\usepackage{amsmath}
\usepackage{amsfonts}
\usepackage{amssymb}
\usepackage{psfig}
\usepackage{makeidx}         
\usepackage{graphicx}        
\usepackage{multicol}        
\usepackage[bottom]{footmisc}

\makeindex             

\begin{document}

\title*{Black hole spin: theory and observation}
\author{M. Middleton}
\institute{M. Middleton, Institute of Astronomy, Madingley Road, Cambridge, \email{mjm@ast.cam.ac.uk}
}
%
%
\maketitle

\abstract*{In the standard paradigm, astrophysical black holes can be described solely by their mass and angular momentum - commonly referred to as `spin'  - resulting from the process of their birth and subsequent growth via accretion. Whilst the mass has a standard Newtonian interpretation, the spin does not, with the effect of non-zero spin leaving an indelible imprint on the space-time closest to the black hole. As a consequence of relativistic frame-dragging, particle orbits are affected both in terms of stability and precession, which impacts on the emission characteristics of accreting black holes both stellar mass in black hole binaries (BHBs) and supermassive in active galactic nuclei (AGN). Over the last 30 years, techniques have been developed that take into account these changes to estimate the spin which can then be used to understand the birth and growth of black holes and potentially the powering of powerful jets. In this chapter we provide a broad overview of both the theoretical effects of spin, the means by which it can be estimated and the results of ongoing campaigns.} 

\abstract{In the standard paradigm, astrophysical black holes can be described solely by their mass and angular momentum - commonly referred to as `spin'  - resulting from the process of their birth and subsequent growth via accretion. Whilst the mass has a standard Newtonian interpretation, the spin does not, with the effect of non-zero spin leaving an indelible imprint on the space-time closest to the black hole. As a consequence of relativistic frame-dragging, particle orbits are affected both in terms of stability and precession, which impacts on the emission characteristics of accreting black holes both stellar mass in black hole binaries (BHBs) and supermassive in active galactic nuclei (AGN). Over the last 30 years, techniques have been developed that take into account these changes to estimate the spin which can then be used to understand the birth and growth of black holes and potentially the powering of powerful jets. In this chapter we provide a broad overview of both the theoretical effects of spin, the means by which it can be estimated and the results of ongoing campaigns.}

\section{Preface}
\label{sec:1}

In the generally accepted model of Einstein's General Relativity (GR - although modified forms of GR cannot yet be ruled out), black holes (BHs) are defined by only their mass, charge and angular momentum, hereafter referred to as spin. In an astrophysical setting the charge will soon neutralise (or else be radiated away during the formation process in accordance with Price's theorem) and so BHs can be entirely defined by only their mass and spin - this is the often-touted `no-hair' theorem. Whilst the effect of mass is relatively benign from the standpoint of observation and theory (generally acting only to scale the energetics e.g. Shakura \& Sunyaev 1973 - and accretion related timescales - e.g. McHardy et al. 2006), the spin has much more to offer in terms of revealing how BHs formed and grew, impacts on how accretion operates in the regime of strong gravity and how the most powerful ejections of material in the Universe may be powered.

What follows should not be viewed as exhaustive but a summary of the state-of-the-art. In section 2 we will review the necessary theory to make sense of the observations and methods that have been applied in attempts to measure BH spin in various systems under various assumptions. In section 3 we will discuss the traditional techniques used to measure the spin in BHBs and AGN and the implications of these measurements (with special attention paid to the launching of astrophysical jets). Sections 4 and 5 discuss new techniques which incorporate the time domain, and in section 6 we conclude with some final remarks about the future prospects for the field.



\section{Useful theory}
\label{sec:2}

Before we discuss how methods to estimate the spin have developed and been applied to observation, it is important to understand how a spinning BH affects the spacetime in which it is embedded as the outcome dictates our approach to studying BHs. Below we present the formalisms which govern the nature of spacetime around such a BH and the behaviour of test particles in close proximity (at large distances this tends towards a Newtonian description); we stress that a deep working knowledge of GR is not necessary to appreciate what follows, with only those formulae considered relevant for our later discussions being presented (though for the more experienced reader we suggest the review of Abramowicz \& Fragile 2013). 

The solution to Einstein's field equations for a spherically symmetric, non-rotating massive body was discovered by Karl Schwarzschild in 1916 for which the metric, which describes the geometry of empty spacetime (a `manifold'), is named, whilst the generalisation to a rotating (uncharged) BH is known as the Kerr metric after Roy Kerr, who discovered the solution in 1963 (see his explanation in Kerr 2008).
 The difference between the two solutions results from the inclusion of the BH angular momentum ({\bf J}) which, as we will see, has a significant impact on the orbit of test particles and the behaviour of infalling material. The metric is usually presented in Boyer-Lindquist co-ordinates (t, r, $\theta$, $\phi$) which can be interpreted as spherical polar coordinates and is related to cartesian co-ordinates via the following standard transforms:
\[\begin{array}{l}
x  = \sqrt{r^{2}+a^{2}}sin\theta cos\phi \\
y  = \sqrt{r^{2}+a^{2}}sin\theta sin\phi \\ 
z  = rcos\theta  
\end{array} \]
Here $a$ is the BH specific angular momentum ($a = \mid${\bf $J$}$\mid/M$) although in literature discussing observation is often expressed in its dimensionless form, $a_{*}$:

\begin{equation}
a_{*} = \frac{|{\bf J}|c}{GM^2}
\end{equation} 
where $M$ is the BH mass in solar units, and $G$ and $c$ are the usual constants. 

In flat (Minkowski) space-time, a line element of size $ds$ is simply given by $ds^{2} = -(cdt)^{2} + dx^{2} + dy^{2} + dz^{2}$, however, in the Kerr space-time (in natural units of G = c = 1) this becomes: 

\begin{multline}
ds^{2} = - \left(1 - \frac{2Mr}{\rho^{2}}\right)dt^{2} + \frac{\rho^2}{\Delta}dr^{2} + \rho^{2}d\theta^{2} + \left(r^{2} + a^{2} + \frac{2Mra^{2}}{\rho^{2}}sin^{2}\theta\right)sin^{2}\theta d\phi^{2}\\ - \frac{4Mra}{\rho^{2}}sin^{2}\theta d\phi dt
\end{multline}
where


\begin{equation}
\rho^{2} = r^{2} + a^{2}cos^{2}\theta 
\end{equation}

\begin{equation}
\Delta = r^{2} - 2Mr + a^{2}
\end{equation}

Using Einstein notation, this can be written in terms of the covariant metric tensor, $g_{\mu\nu}$, where $ds^{2} = g_{\mu\nu}dx^{\mu}dx^{\nu}$, and  $g_{\mu\nu}$ in vector form is:

 \[ \left( \begin{array}{cccc}
g_{tt} & 0 & 0 & g_{t\phi}\\
0 & g_{rr} & 0 & 0\\
0 & 0 & g_{\theta\theta} & 0\\
g_{t\phi} & 0 & 0 & g_{\phi\phi}
\end{array} \right)\]


By inspection, the components of the metric in equation 2 are then:

\begin{equation}
g_{tt} = - \left(1 - \frac{2Mr}{\rho^2}\right) 
\end{equation}

\begin{equation}
g_{t\phi} = \frac{-4Mr}{\rho^2}asin^{2}\theta
\end{equation}

\begin{equation}
g_{\phi\phi} = \left(r^{2} + a^{2} + \frac{2Mra^{2}}{\rho^{2}}sin^{2}\theta\right)sin^{2}\theta
\end{equation}

\begin{equation}
g_{rr} = \frac{\rho^{2}}{\Delta}
\end{equation}

\begin{equation}
g_{\theta\theta} = \rho^{2}
\end{equation}
and setting $a$ = 0 returns the metric for a non-spinning, Schwarzschild BH. 


Whist the above may seem mathematically daunting, the components of the metric play a vital role in allowing us to predict the effect of non-zero spin. In the Schwarzschild metric there are two important radii to consider, the first is the position of the static surface (also called a `null hypersurface') 
of the event horizon, where an observer in a distant reference frame would observe a body, travelling at the speed of light radially away from the BH to be stationary. In this case (i.e. for a non-spinning BH), the event horizon is located at the Schwarzschild radius (which can be easily derived in the Newtonian case with the escape velocity set equal to the speed of light), $R_{s} = 2GM/c^{2}$ (= 2R$_{g}$ or 2M in natural units). The second important radius we need to consider is the position of the marginally stable orbit more commonly referred to as the innermost stable circular orbit (ISCO), inside of which stable orbits are not possible and any accreted material takes a laminar plunge to the event horizon on a dynamical timescale. The position of the ISCO can be found from considering the radial equation of geodesic motion using the proper time (i.e. that measured by a clock at rest) $\tau$:

\begin{equation}
\frac{d^{2}r}{d\tau^{2}} + \frac{M}{r^{2}} - \left(1 - 3M/r\right)u_{\phi}^{2}/r^{3} = 0
\end{equation}
where $u_{\phi}$ is the specific angular momentum. If we consider circular orbits, $d^{2}r/d\tau^{2}$ = 0 so $u_{\phi}^{2} = Mr^{2}/(r-3M)$, and the specific angular momentum has a minimum at r = 6M. Physically this is distinct from the Newtonian case where angular momentum decreases monotonically down to the central object. At radii within the ISCO, circular orbits are no longer possible (see the discussion in Abramowicz \& Fragile 2013) and the material undergoes radial free-fall to the event horizon and the singularity beyond.


Non-zero values of the spin indicate a Kerr BH, with a positive value corresponding to the BH rotating in the same direction as the orbiting particles around it, i.e. prograde. Conversely, negative spin values indicate that the orbits are oriented in the opposite direction, i.e. retrograde to the BH spin. Based on the third law of BH thermodynamics which states that a BH cannot have zero surface gravity (Bardeen, Carter \& Hawking 1973), the spin must have natural limiting values of -1 and 1 (at which point the surface gravity is zero). An additional constraint arises from consideration of the Kerr solution and by setting equation 4 equal to zero:  $\Delta = r^{2} - 2Mr + a^{2} = 0$ which is a coordinate singularity in equation 2. It can be readily seen that there are no real solutions when $a^{2} > M^{2}$ which implies that in such a case there is no horizon and no BH, leading to a `naked singularity' which is forbidden (due to paradoxes); so once again we find a limiting value of $|a| < M$ or (as $a_{*} = a/M$), $|a_{*}| < 1$ as before.


Assuming at the point of BH formation the spin is less than maximal (in the case of stellar mass BHs this is eminently sensible given that angular momentum may be lost in the supernova explosion), accretion of matter in the prograde direction will both increase the mass and `spin-up' the BH, analogous to the situation of accreting neutron stars (e.g. Bisnovatyi-Kogan \& Komberg 1974). Considering only the accretion of matter, the BH's growth follows Bardeen's law (Bardeen, Carter \& Hawking 1973), however, as discussed in Thorne (1974), the effect of radiation is important. If one were to ignore the radiation from the disc then in principle the spin could reach a limiting value of 1, however, as pointed out by Bardeen, Carter \& Hawking (1973), the capture cross section for photons with oppositely aligned momentum is larger than when aligned and so these photons will act to buffer against the spin reaching unity. More accurately, above $a_{*}$ = 0.90, radiation effects {\it cannot} be ignored and lead to a deviation in evolution away from Bardeen's law such that the limiting, `maximal' value is reached at $a_{*}$ = 0.998 for prograde spin or -0.998 in the retrograde case (and changes only very little depending on the nature of the illumination: Thorne 1974). It is important to note that this maximal value does not account for the effect of torques which are expected to result from magnetic fields threading the plunging region which may act to reduce the maximum spin that can be achieved (Gammie, Shapiro \& McKinney 2004).

It is important to note that, whilst accretion must inevitably change the spin (unless maximal already), we cannot yet observe this on any human timescale. From Bardeen's law it can be seen that it would take more than the mass of the BH itself to be accreted to change the spin from 0 to 1; given typical mass transfer rates in BHBs via Roche Lobe overflow of 10$^{-6}-10^{-7}$ M$_{\odot}$/year and outburst duty cycles of $\sim$1\% (e.g. Fragos et al. 2008), it is clear that we will have to wait around the lifetime of the binary itself ($\sim$ billions of years) to see such a substantial change in the spin of stellar mass BHs (and equally for supermassive BHs - SMBHs - in AGN where the duty cycle is thought to be similar but mass loss via winds could be substantial). Smaller changes are however possible on smaller fractions of the binary's lifetime, depending on the starting mass and spin of the black hole (see Fragos \& McClintock 2015), however, for changes in the spin that occur on observable timescales, we presently - and for the foreseeable future - will lack the ability to detect them via the methods we will discuss in the forthcoming sections of this chapter.


%
%


\subsection{Frame-dragging}

A key consequence of having a spinning BH - in terms of observational implications - is the concept of relativistic frame-dragging. As a result of the BH's non-zero angular momentum, space-time moves (is frame-dragged) in the direction of the spin in it's vicinity, thereby imparting energy to an orbiting test particle. This can be seen directly when we consider an observer, with (contravariant) four velocity $u^{\mu}$ (= $dx^{\mu}/d{\tau}$) who falls into the BH with zero angular momentum or L = $u_{\phi}$ = 0. This is the usual definition of a zero angular momentum observer (ZAMO).  The contravariant component of the velocity is non-zero (except as r $\rightarrow$ 0) so $u^{\phi}  = g^{\phi t}u_{t} \neq 0$. The angular velocity of the ZAMO is: 

\begin{equation}
\Omega = \frac{d\phi}{dt} = \frac{\frac{d\phi}{d\tau}}{\frac{dt}{d\tau}} = \frac{u^{\phi}}{u^{t}} \neq 0
\end{equation}

$\Omega$ can be computed from $u_{\phi} = 0 = g_{\phi \phi}u^{\phi} + g_{\phi t}u^{t}$ which gives:

\begin{equation}
\Omega = \frac{u^{\phi}}{u^{t}} = -\frac{g_{\phi t}}{g_{\phi\phi}}
\end{equation}

Substituting equations 5 \& 6 for the components of the metric leads to:

\begin{equation}
\Omega = \frac{2Mar}{(r^{2} + a^{2})^{2} - a^{2}\Delta sin^{2}\theta}
\end{equation}
By substituting equation 4, we can see that $(r^{2} + a^{2})^{2} > a^{2}(r^{2} - 2Mr + a^{2})sin^{2}\theta$ and so $\Omega/Ma > 0$. Therefore, as a result of the non-zero BH spin, a ZAMO is forced to co-rotate (frame-dragged) in the direction of its rotation (as the angular velocity has the same sign as the angular momentum). 

\begin{figure}[t]
\sidecaption
\includegraphics[scale=.55]{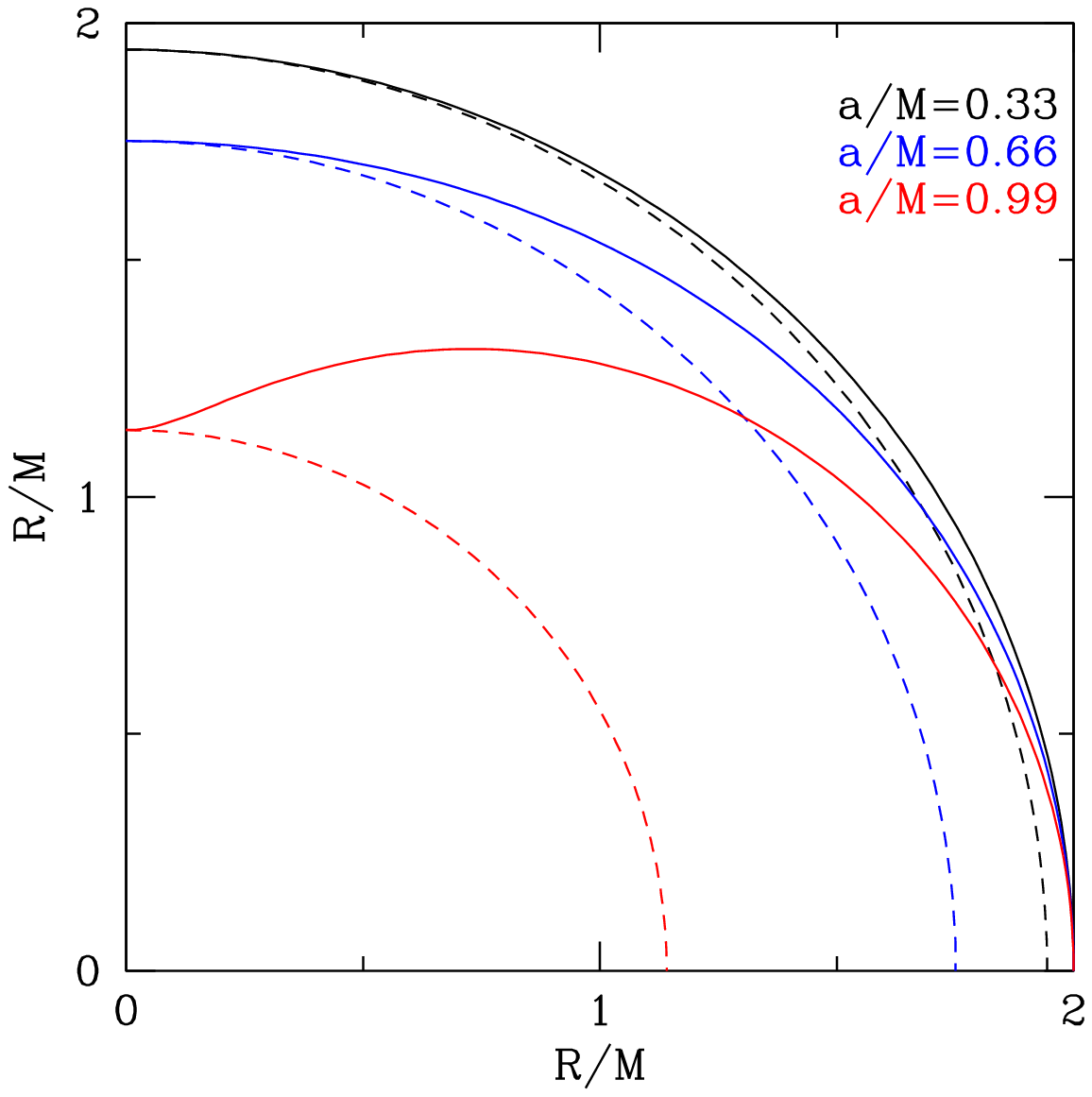}
\caption{2D positions of the event horizon (equation 14: dashed line) and ergosphere (equation 16: solid line) as a function of BH spin (across polar coordinate space - where the x-axis is the radial distance from the BH). The hypersurfaces meet at the poles (co-latitude of 0 degrees) with the ergosphere `pinched' down as the spin increases.
}
\label{fig:2}       
\end{figure}


As a consequence of frame-dragging, whilst the Schwarzschild metric is only singular (in Boyer-Lindquist coordinates) at the static surface of the event horizon, solutions are singular across two null hypersurfaces in the Kerr metric. The new position of the event horizon is found where $g_{rr}$, the radial component of the metric (equation 2) tends to infinity. By setting 1/$g_{rr}$ = 0, we can see that this is the same as the solution to the coordinate singularity at $\Delta = 0 = r^{2} - 2Mr + a^{2}$. Solving the quadratic leads to the solution for the radius of the horizon: 

\begin{equation}
r_{H} = M \pm \sqrt{M^{2} - a^{2}}
\end{equation}


The positive solution defines the event horizon (as radii below this are forced to travel faster than the speed of light). As we can see, for non-zero spin, the position of the event horizon can be within the Schwarzschild radius. The second hypersurface occurs when $g_{tt}$ changes sign or, from equations 4 and 5:

\begin{equation}
g_{tt} = 0 = -\left(1 - \frac{2Mr}{r^{2} + a^{2}cos^{2}\theta}\right)
\rightarrow 0 = r^{2} - 2Mr + a^{2}cos^{2}\theta
\end{equation} 
which has the solutions:

\begin{equation}
r_{E} = M \pm \sqrt{M^{2} - a^{2}cos^{2}\theta}
\end{equation}
 
The surface at $r_{E+}$ is referred to as the `ergosphere' and the region between this and the event horizon the ergoregion. It is easy to see that $\sqrt{M^{2} - a^{2}cos^{2}\theta} > \sqrt{M^{2} - a^{2}}$ for all co-latitudes ($\theta$) except at the poles where the position of the event horizon and ergosphere meet (see Figure 1). 

As  $r_{E+} > r_{H+}$ (when $\theta \neq 0$ and $\theta \neq \pi$), an observer within the ergosphere can still be in causal contact with the outside Universe; this is plotted in Figure 2. Within the ergosphere, it is not possible for a physical observer to remain at rest and, from calculating the effect of orbits within this region, it was discovered that negative energy (retrograde) trajectories/orbits are possible (see e.g. Penrose 1969; Bardeen et al. 1972). Should an orbiting body fragment within the ergosphere, then the total energy of those fragments not induced into negative energy orbits will be greater, having effectively tapped the energy (angular momentum) of the BH. This tapping of the BH's spin is called the Penrose effect (Penrose 1969) and the magnetic field analog, the Blandford-Znajek effect (Blandford \& Znajek 1977), which we shall discuss in section 3.5 in relation to powering `superluminal' ejections.

\begin{figure}[h]
\sidecaption
\includegraphics[scale=.6]{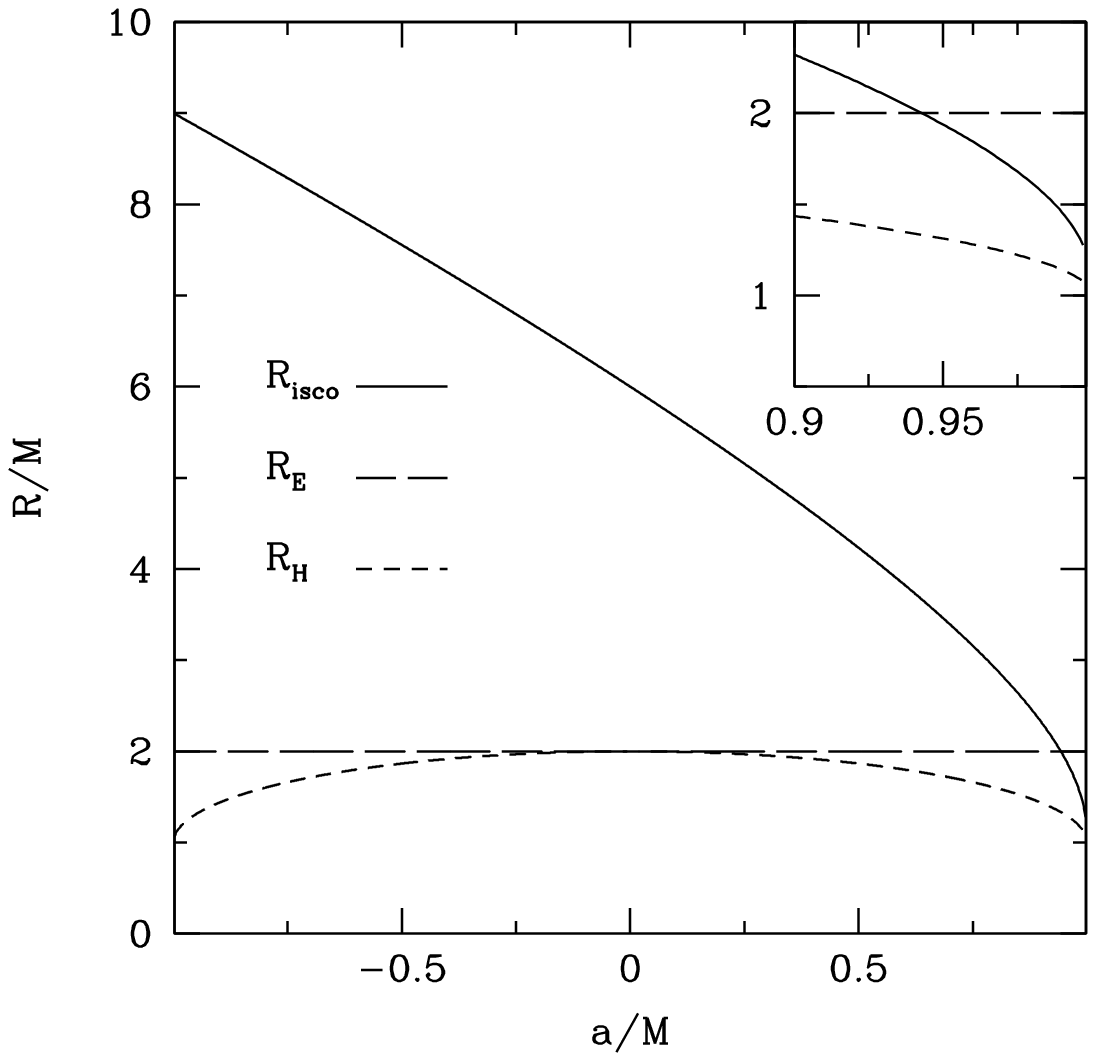}
\caption{Position of the respective hypersurfaces in the Kerr metric as a function of BH spin (equations 14, 16 \& 17), R$_{\rm H}$, the event horizon,  R$_{\rm E}$, the ergosphere (shown for $\theta$ = $\pi/2$, i.e. in the plane of an accretion disc) and R$_{\rm isco}$, the innermost stable circular orbit. The top right inset shows an enlarged version covering the highest spins where R$_{\rm isco}$ sits {\it inside} the ergosphere.}
\label{fig:2}       
\end{figure}

As stable orbits are possible closer to the BH for prograde spin (Bardeen et al. 1972) frame-dragging changes the position of the ISCO which is a well-defined, monotonic function of a/M (Bardeen et al. 1972) and is shown in Figure 2: 

\begin{equation}
r_{isco} =  M\left[3 + Z_{2} \mp [\left(3 - Z_{1}\right)\left(3 + Z_{1} + 2Z_{2}\right)]^{1/2}\right]
\end{equation}
where

\begin{equation}
Z_{1} = 1 + \left(1 - a^{2}/M^{2}\right)^{1/3}[\left(1 + a/M\right)^{1/3} + \left(1 - a/M\right)^{1/3}]
\end{equation}

\begin{equation}
Z_{2} = \left(3a^{2}/M^{2} + Z_{1}^{2}\right)^{1/2}
\end{equation}
where $\mp$ is used, the top sign refers to a treatment where the spin is prograde and the bottom sign to where the spin is retrograde.

As can be seen from Figure 2, for $a > $0, the position of the ISCO lies within that for a Schwarzschild BH (6M), reaching a minimum at 1.235M (for $a_{*}$ = 0.998) and is pushed further out in the case of $a < $0 (retrograde spin) towards a maximum of 9M. As we shall see in the following sections, this correspondence between the spin and location of the ISCO allows us to construct models to estimate the spin from observation.

It is worth noting as a point of general interest that, in addition to the changes to the positions of the hypersurfaces discussed above, the singularity itself can no longer be point-like but must take the form of a ring (we will not discuss the effects of ring singularities further but point the interested reader to Burko \& Ori 1997). Additionally, although we have discussed the effect of frame-dragging on test particles, electromagnetic fields (either generated in the flow via the MRI: Balbus \& Hawley 1991 or in the local environment to the BH) will also be affected and can lead to reconnection events and particle acceleration (e.g. Karas, Kop{\'a}{\v c}ek \& Kunneriath 2012).


\subsubsection{Relativistic precession}

There are two further implications of a spinning BH, resulting from the effect of frame dragging; Lense-Thirring precession and the Bardeen-Peterson effect, both of which may provide important observational diagnostics of the spin and region of strong gravity close to the BH. 

Lense-Thirring precession (also called the Lense-Thirring effect: Lense \& Thirring 1918) describes the behaviour of orbiting and vertically displaced motion in proximity to a rotating massive body. For this reason it is relevant not only for Kerr BHs but more generally relevant for satellites of astrophysical bodies such as stars and planets. 
Due to frame-dragging, the orbital motion undergoes precession leading to epicyclic oscillations about periapsis (position of closest approach) and the ecliptic as long as the orbit is vertically misaligned with the rotation axis of the rotating body.


Should the accretion disc be misaligned with the BH spin axis (for instance due to a supernova kick, see Brandt \& Podsiadlowski 1995) then the precession of orbits due to the Lense-Thirring effect produces a torque. If this torque is larger than the viscous torques in the disc, a fluid (non-solid body) inner disc will align perpendicular to the BH spin axis whilst beyond the warp radius the disc aligns with the binary orbit; this is the Bardeen-Petterson effect (Bardeen \& Petterson 1975). Should the torques not be dissipated (though see discussions by Armitage \& Natarajan 1999; Markovic \& Lamb 1998), the whole of the inner disc can therefore precess leading to important effects on the emergent spectrum and jets (see section 4). Such precession may lead to clear signatures in the time domain as well as the energy domain (Ingram \& Done 2012) and as we will discuss later, tests for the BH spin. 


The manner in which the disc warp created by the Bardeen-Peterson effect propagates depends on the nature of the disc: if it is thin or the viscosity high then the warp diffuses (due to viscous torques as discussed by Pringle 1992), whilst if the disc is thick or viscosity low, the warp propagates as a wave (Papaloizou \& Lin 1995). King et al. (2005) discuss the general case (for both viscosity cases and for the full range of disc tilts and misalignment) of how the torque between the BH and disc as a result of precession can lead to co-alignment or counter-alignment of the BH/disc system. The authors conclude that alignment depends on the detailed properties of the disc, namely how the warp is propagated, although, in general, on short timescales it is possible that the disc tries to misalign with the hole (essentially spinning the hole down), whilst on long timescales there is a tendency towards co-alignment (and spin-up). Making comparisons to observation, Maccarone (2002) report that the jet angle in the BHBs, GRO J1655-40 and SAX J 1819-2525, are misaligned with respect to (i.e. not perpendicular to) the binary orbit. Assuming the jet angle is tied to the inner disc and the spin of the BH, then this may indicate that the mis-alignment resulted from the formation process (e.g. Brandt \& Podsiadlowski 1995) and the time required for alignment is potentially a significant fraction of the binary lifetime (Martin, Tout \& Pringle 2008; Steiner \& McClintock 2012). King et al. (2005) note that in such systems the angular momentum of the BH is much larger than that of the disc and so the crucial timescale is that on which tidal forces can transfer angular momentum from the binary orbit to the disc, on timescales shorter than this, counter-aligned discs may be possible. 

In the case of AGN, the picture is less clear as the timescales for accretion driven changes (and alignment) are considerably longer (typically scaling with mass) whilst the means by which material reaches the inner sub-pc disc is still debated. Should material fall through the galactic disc then the angular momentum is expected to be in a single direction and the BH's spin axis should appear aligned with the host galaxy's stellar disc (assuming that the growth is driven by accretion rather than via BH-BH mergers). Instead, should material with a range of angular momenta be accreted (e.g. via condensed  filaments: Nayakshin, Power \& King 2012) - a situation often referred to as ``chaotic" accretion - then it is possible that the disc-BH system will be initially mis-aligned with respect to the inflowing material (which is assumed to be misaligned with the host galaxy's stellar disc) and then co-align - as a consequence, the sub-pc system does not need to be aligned with the galactic plane. Observations of jets (see Hopkins et al. 2012) indicate that many AGN-galaxy systems do indeed appear misaligned which likely points towards a recent chaotic accretion history.


\section{Observational  tests of spin I - the energy spectral domain}

The effect of frame dragging and the change in the position of the ISCO with spin (Figure 2) has led to the development of methods by which the BH spin can be estimated. We purposely use the word estimated here to signify the uncertainty inherent in characterising such an intrinsic yet complex property, relying on models themselves based on caveats and assumptions. However, this should {\it not} be read as a criticism of efforts both past and ongoing to better estimate and constrain the spin, rather that the reliability of a chosen method should be evaluated against the backdrop of systematic uncertainties. 

\subsection{Modelling the Continuum (disc) spectrum}

As mentioned elsewhere in this compilation, transient (predominantly low-mass companion) BHBs, undergo outburst cycles regulated by disc-instabilities (e.g. Lasota et al. 2001) evolving in brightness and spectral shape (see McClintock \& Remillard 2006 and Done, Gierli{\'n}ski\& Kubota 2007 for reviews). Towards the peak of the outburst the spectrum becomes increasingly dominated by emission originating from the accretion disc. Under the assumption that the inner disc radius (R$_{in}$) sits at the ISCO (as seen in GRMHD simulations: Shafee, Narayan \& McClintock 2008; Penna et al. 2010  - and see Zhu et al. 2012 for the effect of the plunging region - and evidenced in the observational study of LMC X-3 by Steiner et al. 2010), we can show that the properties (namely the temperature and luminosity) are related to the spin through the following formulae (for a rigourous discussion we point the reader to Frank, King \& Raine 1985). We point out that the following derivations are meant only to illustrate the case using the simplest, non-relativistic treatment (we discuss the relativistic disc modelling in section 3.1.1). The presence of viscous torques on the differential (Keplerian) orbits leads to dissipation of mechanical energy with the torque defined as:

\begin{equation}
t_{\phi}(R) = 2\pi Rv\Sigma R^{2}
\end{equation}
where $v$ is the kinematic viscosity 
, $\Sigma$ is the surface density and $\Omega'$ is the radial gradient of angular momentum (d$\Omega$/dr).

Although $v$ becomes unimportant in terms of calculating the emission profile (as we shall soon see), its form is relatively important historically as it can also be parameterised as:

\begin{equation}
v = \alpha c_{s}H
\end{equation}
where $c_{s}$ is the sound speed, H is the height of the disc and $\alpha$ is the viscosity parameter that underpins the $\alpha$-prescription of Shakura \& Sunyaev (1973). This formula assumes that turbulence drives the viscosity and results from a consideration of the typical size of a turbulent eddy (which must be less than the disc scale-height, H/R) and the assumption that the turbulent velocity is not supersonic. As a consequence we would expect $\alpha < 1$. As Frank, King \& Raine (1985) point out, this is not a physical statement as the true nature of the viscosity is unknown (although they also point out that magnetic stress would also lead to $\alpha < 1$).

Irrespective of the nature of the viscosity, the amount of mechanical heat loss is given by t$_{\phi}$(R)$\Omega'$dR and is dissipated across both sides of the disc (2$\times$2$\pi$RdR), giving a heat loss per unit area ($D(R)$) of:

\begin{equation}
D(R) = \frac{t_{\phi}\Omega'}{4\pi R} = \frac{v\Sigma}{2}(R\Omega')^2
\end{equation} 

Setting $\Omega$ to be Keplerian (i.e. differential rotation: $\Omega_{k} = (GM/R^{3})^{1/2}$), we find:

\begin{equation}
D(R) =  \frac{9}{8}v\Sigma \frac{GM}{R^3}
\end{equation} 

From conservation of mass and angular momentum, and assuming zero torque at the innermost edge of the disc (i.e. $\Omega'$ = 0 at R = $R_{in}$; see Krolik et al. 1999; Gammie 1999 and Balbus 2012 for issues associated with this assumption) it can be seen that:


\begin{equation}
v\Sigma =  \frac{\dot{M}}{3\pi}\left[1-\left(\frac{R_{in}}{R}\right)^{1/2}\right]
\end{equation} 

Combining equations 23 and 24 leads to the formula for the dissipation of energy per unit area more commonly seen in literature:

\begin{equation}
D(R) = \frac{3GM\dot{M}}{8\pi R^{3}}\left[1-\left(\frac{R_{in}}{R}\right)^{1/2}\right]
\end{equation}
which importantly demonstrates that the heating is {\it independent} of viscosity (either related to the sum of radiation and gas pressure in the disc: Shakura \& Sunyaev 1973, or via magnetic stresses caused by the MRI: Balbus \& Hawley 1991). Assuming the disc to be fully optically thick to the emergent thermal radiation, we should expect local emission (i.e. at each radius) to be a blackbody (Planck distribution) with a peak, effective temperature, $T_{eff}$ according to Stefan Boltzmann's law:

\begin{equation}
D(R) = \sigma_{SB} T_{eff}^4
\end{equation}
or 

\begin{equation}
T_{eff} = \left\{\frac{3GM\dot{M}}{8\pi R^{3}\sigma_{SB}}\left[1-\left(\frac{R_{in}}{R}\right)^{1/2}\right]\right\}^{1/4}
\end{equation}

Thus, in this non-relativistic approximation, the temperature of the thermal emission is in principle related to the position of the inner radius and is therefore a diagnostic of the spin. 

The luminosity that emerges as a result of the process of accretion through the disc can be approximated by:

\begin{equation}
L = \frac{GM\dot{M}}{2R}
\end{equation}
where R is the position of the inner edge of the flow and the factor of 2 in the denominator results from the virialisation of the system (i.e. half of the potential energy is radiated whilst the remaining half is converted into kinetic energy and lost to the BH). This luminosity can also be parameterised as the conversion of rest mass to energy given by:

\begin{equation}
L = \eta \dot{M}c^{2}
\end{equation}
where $\eta$ is the radiative efficiency. By equating the two formulae above it is clear that, in the simplest picture, the radiative efficiency is a function of the position of the inner edge and therefore the spin, ranging from $\sim$8-40\% for zero through to maximal (prograde) spin. This is the simplified Newtonian case and is a good approximation to the actual efficiency as a function of spin which goes as:

\begin{equation}
\eta = 1 - \left(R_{ISCO} - 2M \pm A_{1}\right)\left(R_{ISCO} - 3M \pm 2A_{1}\right)^{-1/2}
\end{equation}
where $A_{1} = a\sqrt{M/R_{ISCO}}$. We note that the above equations assume that the mass falling onto the BH is only converted into radiation. This is demonstrably not the case as powerful winds and jets are ubiquitous to accretion flows, however, this remains an important illustrative point and a useful theoretical framework for discussing BH accretion discs.

In practice the emission spectrum from the accretion disc is a convolution of the thermal emission from all radii or a `multicolour' disc blackbody (e.g. Mitsuda et al. 1984; Makishima et al. 1986). In addition to this deviation from a blackbody, a further complication arises due to the effect of opacity which determines how deep into the disc atmosphere we observe, i.e. the position of an `effective photosphere', $\tau_{eff}$. The two `competing' forms of opacity are electron scattering ($\kappa_{T}$) and absorption via both free-free ($\kappa_{ff}$) and via metal-edges/bound-free transitions ($\kappa_{bf}$). Whilst $\kappa_{T}$ is independent of temperature and density, both forms of absorption opacity scale as $\rho*T^{-7/2}$ (Kramer's law). Thus, the position of 
$\tau_{eff}\approx\tau_{T}\sqrt{\kappa_{abs}/\kappa_{T}} = 1$ is a function of temperature/frequency where $\kappa_{abs}$ is the sum of the contributions to the absorption opacity. At higher frequencies, we can see further into the disc as the absorption opacity is lower; as there is a negative, vertical temperature gradient through the disc, when $\tau_{eff}=1$ is further into the disc, we observe a larger offset in temperature compared to the surface. Such effects lead to the requirement of a colour correction/spectral hardening factor, $f_{col}$ where the {\it observed} temperature of a blackbody at a given radius, $T_{col} = f_{col}*T_{eff}$. Such that the intensity at a given frequency ($I_{\rm\nu}$):



\begin{equation}
I_{\nu} = \frac{1}{f_{col}^4}B_{\nu}(T_{col})
\end{equation}
and $f_{col}$ can be roughly parameterised by the ratio (i.e. relative importance) of the competing opacities in the disc:

\begin{equation}
f_{col} \sim \left(\frac{\kappa_{tot}}{\kappa_{abs}}\right)^{1/4}
\end{equation}
where $\kappa_{tot} = \kappa_{abs} + \kappa_{T}$ and $f_{col}$ reaches saturation (Davis, Done \& Blaes 2006) at:

\begin{equation}
f_{col} \sim \left(72~keV/T_{eff}\right)^{1/9}
\end{equation}

The simplest and most widely adopted disc model, {\sc diskbb} (for use in the spectral fitting package {\sc xspec}: Arnaud 1996 or ISIS: Houck \& Denicola 2000), which does not include $f_{col}$ but is commonly used to describe the thermal emission seen in BHBs and neutron star binaries and can be used to provide a crude estimate of $R_{in}$ and therefore the BH spin. 




\subsubsection{Beyond the simple picture}

A more accurate picture of accretion in the framework of GR was developed by Novikov \& Thorne (1973) and Page \& Thorne (1974), assuming a razor-thin disc and zero-torque inner boundary condition (and can be seen as the relativistic analog to the Shakura-Sunyaev disc), more commonly referred to as the General Relativistic accretion disc (GRAD) model. Building upon this relativistic framework, models are now available that include the full `suite' of relativistic corrections (Doppler boosting and gravitational redshift), the effect of returning radiation and importantly non-zero inner-boundary conditions (i.e. $t_{\phi}(R_{in}) \neq$ 0). This last point is hotly debated as magnetic fields crossing the ISCO may connect the disc to the BH or plunging region and thereby provide a torque (see discussions by van Putten 1999; Paczy\'{n}ski 2000; Armitage, Reynolds \& Chiang 2001; Hawley \& Krolik 2002; Afshordi \& Paczy\'{n}ski 2003; Li 2003). One of the most widely used of the GRAD models is {\sc kerrbb} (Li et al. 2005) which includes a grid of spectra created via `ray-tracing' in the Kerr metric. The method of ray-tracing is a well established and reliable means of mapping photon paths in a given metric which can be seen as a way to effectively visualise emission from the accretion flow (e.g. Cunningham  \& Bardeen 1973; Cunningham 1975; Rauch \& Blandford 1994; Fanton et al. 1997; $\breve{C}$ade$\breve{z}$ et al. 1998; M$\ddot{u}$ller \& Camenzind 2004; Schnittman \& Bertschinger 2004; Dexter \& Agol 2009). In practice, the disc `image' seen by an observer in some observer-system geometry is broken into a number of small elements and photon paths are traced back to the disc. By assuming a local flux-density profile at each location in the disc and by incorporating relativistic effects (Doppler boosting and gravitational redshift) the final spectrum can be reconstructed by summing over the disc elements the paths intercept. In addition to direct illumination, ray-tracing also allows photons to return to the disc from the far side due to gravitational light-bending, leading to a change in the locally emitted flux. The grids in the {\sc kerrbb} model allow for a range of spin, inclination and BH mass whilst assuming a standard disc structure with constant $f_{col}$ (= 1.7, although this value can be changed in the model) and allows for limb-darkening (e.g. Svoboda et al. 2009). 



In determining $f_{col}$ in the above models, bound-free absorption has been ignored, however it can dominate over free-free opacity and lead to changes in $f_{col}$ with an increased likelihood that photons are instead `destroyed' rather than propagated and scattered. At the time of writing, the only disc model which incorporates metal edges is {\sc bhspec} (Davis et al. 2005; Davis \& Hubeny 2006). This model describes a GRAD (Novikov \& Thorne 1973) but {\it unlike} {\sc kerrbb}, calculates the disc spectrum by including a relativistic transfer function in place of ray-tracing. The transfer function provides the integration kernel in calculating the disc emission and contains information regarding the Doppler boost due to rotation and gravitational light-bending (see Cunningham 1975, 1976; Laor 1991; Speith, Riffert \& Ruder 1995; Agol 1997; Agol \& Krolik 2000; Dov$\breve{c}$iak, Karas \& Yaqoob 2004).  

Davis et al. (2005) create the {\sc bhspec} model tables following the methods laid out in Hubeny et al. (2000, 2001 and references therein) adopting the {\sc tlusty} stellar atmosphere code (Hubeny \& Lanz 1995) to solve the equations for the vertical structure and angular dependence of the radiative transfer. In addition to including bound-free opacities (assuming ground state populations) the model fully accounts for Comptonisation of the escaping radiation. The authors find that bound-free absorption does indeed play an important role in determining $f_{col}$ with typical values between 1.5-1.6, (lower than found by Shimura \& Takahara 1995 and Merloni et al. 2000) with the actual value depending on the mass accretion rate and spin (and weakly on the adopted value of the $\alpha$ parameter or stress prescription). As with {\sc kerrbb}, {\sc bhspec} accounts for the angular dependence of limb-darkening.  

The hybrid code, {\sc kerrbb2} combines the ability of {\sc bhspec} to self-consistently determine spectral hardening values with the ability of {\sc kerrbb} to account for returning radiation. To do this, tables of spectral hardening values are pre-computed (publicly released for {\it RXTE} only) by fitting spectra generated with {\sc bhspec} with {\sc kerrbb} at fixed sets of input values (see McClintock et al. 2006 for an example of its use). 

In the case of the above models it is important to have reliable estimates of the system parameters, notably the inclination which sets the amount of limb darkening and Doppler boosting (and visible area from which the flux is emitted), the BH mass which shifts the peak temperature, sets the Eddington limit (and in extreme cases can influence $f_{col}$ through changing the density) and the distance which sets the luminosity (and therefore Eddington ratio). As the luminosity scales as the inverse square of the distance, the resulting spin values are highly sensitive to this parameter and obtaining an accurate distance measurement (for instance via radio parallax measurements: see Miller-Jones 2014) is critical. 


As we will discuss in section 3.3, the above models have been widely utilised in estimating the spins for BHBs. However, whilst lauding their successes it is also important to consider the limitations of such models. There are a number of key assumptions that go into both {\sc kerrbb} and {\sc bhspec} including the assumption that the discs are steady-state (i.e. time independent) and that the flows are radiatively efficient. The first assumption breaks down should winds be driven from the innermost disc by thermal, MHD processes or radiative pressure (e.g. Ponti et al. 2012) as the mass accretion rate is then a function of time and radius. The second assumption is likely to be invalid at very low mass accretion rates where the flow is low density and hot material is carried into the BH before it can radiate, i.e. advected (Narayan \& Yi 1994; 1995); such flows are referred to as advection dominated accretion flows (ADAFs) or radiatively inefficient accretion flows (RIAFs). Such ADAFs/RIAFs also appear at very high mass accretion rates where the scale height of the disc is very large and so photons are effectively trapped as the diffusion timescale is longer than the viscous infall timescale (Ambramowicz et al. 1988). The effect of advection at high accretion rates is to stabilise the disc (i.e. removes heat from the flow) with a change in the emission profile from $T \propto R^{-3/4}$ (see equation 27) to $T \propto R^{-1/2}$. Non-relativistic models (with multicolour discs) with a free emissivity profile are available (e.g. {\sc diskpbb}: Mineshige et al. 1994, Hirano et al. 1995, Watarai et al. 2000, Kubota \& Makishima 2004, Kubota et al. 2005) however, until recently these did not include the complex calculations necessary to describe a physically motivated disc atmosphere and relativistic transfer/ray-tracing important in the Kerr metric. 

The model {\sc slimbb} (Straub et al. 2011; Sadowski et al. 2011) utilises ray-tracing and the radial and vertical profile solutions given in Sadowski et al. (2009) and Sadowski et al. (2011) . This model once again does not account for mass loss in a wind but accounts for three key components in high mass accretion rate ADAFs, the radial advection of heat and subsequent change in the emissivity, the position of the inner edge (which can move from the marginally stable to marginally bound orbit: Abramowicz et al. 1988) and the location of the effective photosphere. A later version of this code, {\sc slimbh} (Straub, Done \& Middleton 2013), incorporates the {\sc tlusty} stellar atmosphere code directly and so is closer in nature to {\sc bhspec}.

Unlike the case for high mass accretion rate ADAFs, the emission from low accretion rate ADAF/RIAFs may arise from
synchrotron cooling in radiatively inefficient jets (Fender, Gallo \& Jonker 2003) or strong outflowing winds (Blandford \& Begelman 1999), Bremmstrahlung emission or Compton scattering in the plasma. Whilst the exact nature of the emission is still debated (although predicted correlations between radio and X-ray luminosity may tend to favour emission from an outflow, e.g. Corbel, Koerding \& Kaaret 2008; Gallo et al. 2014), it is clear that the optically thick disc is not present and therefore such accretion states are not presently used to diagnose the spin.

There is also a more fundamental assumption that goes into models that derive from the Novikov \& Thorne (1973) prescription: real accretion discs will have a finite thickness and will not behave as if razor thin. Paczy\'{n}ski (2000) and Afshordi \& Paczy\'{n}ski (2003) argue for a monotonically decreasing deviation with decreasing scale-height for small $\alpha$. This assertion was later confirmed by calculations (Shafee, Narayan \& McClintock 2008) and so in these limits, the GRAD models should be reliable. However, this does not account for the presence of magnetic fields in the disc which are expected to be generated via a dynamo effect and give rise to magnetic stresses and angular momentum transport (Balbus \& Hawley 1991). Recent 3D GRMHD simulations (Shafee et al. 2008; Noble, Krolik \& Hawley 2009, 2010; Penna et al. 2010) estimate that both the luminosity and stress in the inner regions differs substantially (by up to 20\%: Noble et al. 2009) to that expected in the Novikov-Thorne prescription. Both Kulkarni et al. (2011) and Zhu et al. (2012) explore how this might affect estimates for the spin derived from the use of codes such as {\sc kerrbb} and {\sc bhspec}. The former obtain the disc flux profile for a series of spin values resulting from the 3D GRMHD thin-disc simulations of Penna et al. (2010), setting $f_{col}$ = 1.7 (as with {\sc kerrbb}), arguing that the extra sophistication of calculating the position of the effective photosphere is unnecessary in this instance. The spectra themselves are then determined via ray-tracing (see the discussion on {\sc kerrbb}) without returning radiation but taking into account limb-darkening. Zhu et al. (2012) perform a similar set of GRHMD simulations but, by including {\sc tlusty} and radiative transfer, also include the distortion to the spectrum as a result of spectral hardening (as described above). Both sets of authors find that, unlike Novikov-Thorne discs, emission can be seen to originate from within the ISCO due to a combination of advection at high accretion rates (as discussed above: Sadowski 2009) and non-zero inner torque, resulting from a finite disc thickness and leading to increased viscous dissipation at radii close to the ISCO. The combination of these two effects leads to the emission peaking at smaller radii than Novikov-Thorne discs giving both a higher peak temperature and larger emitted flux (and which extends to outside of the plunging region: Zhu et al. 2012). By fitting the simulated spectra with {\sc kerrbb} (Kulkarni et al. 2011) and {\sc bhspec} (Zhu et al. 2012), the spin is found to be systematically overestimated as a result of the increased disc brightness. Crucially, the typical error resulting from use of the Novikov-Thorne profile is far less than the errors associated with the system parameters of mass, inclination and distance (see Gou et al. 2009; Steiner et al. 2010); the spin values derived from GRAD based models are therefore reliable where the errors on the system parameters dominate.


As an addendum, it is also possible that the disc emission profile
differs dramatically from all of those mentioned thus
far. Novikov-Thorne (and Shakura-Sunyaev) discs are both thermally
(Shakura \& Sunyaev 1976) and viscously (Lightman \& Eardley 1974)
unstable whilst MHD simulations carried out by Hirose, Krolik \& Blaes
(2009) find that small patches of the disc can be thermally
stable yet viscously unstable. As a result (and motivated by both
spectra and variability arguments) Dextor \& Agol (2011) proposed a
toy model for an inhomogeneous disc (ID) which consists of zones which
undergo independent fluctuations driven by radiation pressure
instabilities and leads to random walks in the effective temperature
at that radius. As Dexter \& Quataert (2012) point out, this model can
explain the spectral and variability properties of the soft states in
BHBs. Importantly, should this model be an accurate depiction of the
disc and its emission, there are implications for our ability to
measure the spin from fits to the continuum. The ID model assumes a
magnitude of temperature fluctuations, $\sigma_{T}$ (not to be
confused with the Thompson cross section); increasing this value leads
to a higher characteristic temperature at the inner edge (see equation
27). As a consequence, deriving the position of the ISCO from the
temperature using the Novikov-Thorne disc profile could be misleading,
resulting in systematic overestimates for the spin (as with the
consideration of a finite thickness disc discussed above). Dexter \& Quateart
(2012) quantify the likely effect, finding that the impact depends on
the model used to account for the hard tail accompanying the disc
emission (McClintock \& Remillard 2006). In practice, they determine that the errors can be far
larger than those resulting from deviation from the Novikov-Thorne
prescription resulting from a finite disc thickness (Kulkarni et
al. 2011) except in the most disc dominated states (with a disc
fraction $\gtrsim$ 0.95 or $\sigma_{T} \lesssim$ 0.15).

\subsection{Modelling the reflection spectrum}

A hard tail of emission ubiquitously accompanies the thermal disc emission in BHBs (and AGN - see e.g. Jin et al. 2012) and must originate by inverse Comptonisation in a corona of thermal/non-thermal plasma in some, as yet, undetermined geometry (e.g. Liang \& Price 1977; Haardt, Maraschi \& Ghisellini 1994). The exact spectral shape (and properties such as variability) of this component are discussed in detail elsewhere in this compilation but can very broadly (though not always accurately) be described by a power-law which breaks at the electron temperature of the thermal particle distribution.  

In all models, the corona producing the power-law emission has a larger scale-height than the disc and as a consequence, the disc will subtend some solid angle to the up-scattered seed photons which will re-illuminate the disc down to a typical scattering surface at $\tau_{eff}$ = 1 (e.g. Guilbert \& Rees 1988; Lightman \& White 1988; George \& Fabian 1991). The resulting `reflection' spectrum is composed of scattered re-emission (i.e. Compton scattering in the surface layers), bound-free edges, bound-bound absorption and emission lines. The three most important components of the spectrum are the Fe K$_{\alpha}$ edge, the Fe K$_{\alpha}$ fluorescent line (s) and the Compton down-scattered hump. The details of these components are discussed in detail in the review of Fabian et al. (2000) with only a brief overview provided here. 

As Fe is cosmically abundant and the fluorescent yield (the probability that a fluorescent line is produced following photoelectric absorption) of neutral elements scales as $Z^{4}$, emission from Fe is expected to be of great importance. Moderate energy (a few keV) X-ray photons produced via inverse Compton scattering in the corona are energetic enough to remove the inner K (1s) shell electron (leading to a sharp photoelectric edge at 7.1~keV). As long as electrons are available in the L (2s) shell, one will drop to fill the K-shell gap and release a photon at either 6.404~keV (K$\alpha_{1}$) or via spin-orbit interaction, a secondary transition at 6.391~keV (K$\alpha_{2}$). The probability of this occurring is only 34\%, whilst the most favoured (66\% probability) outcome is Auger de-excitation where the production of a photon via the L$\rightarrow$K shell transition is absorbed by a bound electron which is subsequently expelled. These energies assume that the Fe is neutral, however, illumination of the disc will act to increase the ionisation state of the reflecting material (e.g. Ross \& Fabian 1993; Ross, Fabian \& Young 1999), parameterised as $\xi = L/nr^{2}$, where $n$ is the electron density and $r$ is the distance from the ionising flux of luminosity $L$ (Tarter et al. 1969). An increase in ionisation state increases the electron binding energy and the energy of the edge and Fe K fluorescent doublet accordingly, however, the lines only emerge significantly above 6.4~keV for Fe XVII and above. The fluorescent yield will also (weakly) depend on ionisation state and when Fe is Li- through to H-like, Auger de-excitation is no longer possible as two L shell electrons are required. Instead photo-electric recombination can lead to line emission (with a high fluorescent yield) at $\sim$6.8~keV.

Although Fe K is the strongest feature of the reflected emission, other metal transitions (e.g. Ni K$_{\alpha}$) also contribute to the overall picture through line emission and absorption edges. Thus the precise details of the reflected emission clearly depend on the elemental abundances and ionisation state of the illuminated material and have been discussed in detail by Matt, Fabian \& Ross (1993; 1996) who consider illumination of a constant density slab. They find four distinct regimes of increasing $\xi$, ranging from `cold' reflection for $\xi <$ 100 ergs/cm/s (where reflection around Fe K resembles that expected from neutral material and the absorption edge is saturated and weak) and terminating at $\xi >$ 5000 ergs/cm/s where there is no absorption edge or line. In between these regimes, the strength of the line is dependent on the number of electrons available in the L shell as described above, with a weak line and moderate edge produced by ionised species of Fe up to Fe XXIII  (where the Auger effect may still take place) and a stronger `hot' line by species above FeXXIII (where the Auger effect no longer takes place and line emission is a result of recombination). In the latter case, the edge appears stronger due to increased flux below the edge as a result of diminishing opacity. One of the most successful models which accounts for illumination onto a semi-infinite slab of optically thick material in the atmosphere of an accretion disc is {\sc reflion} and its later incarnation, {\sc reflionx} (Ross, Fabian \& Young 1999; Ross \& Fabian 2005). These models, based on the work of Ross, Weaver \& McCray (1978) and Ross (1979), fully incorporate both the radiative transfer of continuum X-rays (using the Fokker-Planck diffusion equation), line emission and Comptonisation (using the modified Kompaneets operator) and thus were an important step-forward from earlier models which, whilst not including line emission or intrinsic emission from inside the gas, were the first to incorporate Green's functions to describe the scattering of photons by electrons in cold gas (e.g. {\sc pexrav}: Magdziarz \& Zdziarski 1995).

The important effects in creating the reflection spectrum discussed so far make no mention of the effect of spin but this has a significant impact on the emergent spectrum for a number of reasons. Principally, the Fe K$_{\alpha}$ emission line (s) originates from illuminated disc radii which rotate in circular Keplerian orbits (although deviations are expected to scale with (H/R)$^{2}$). The observed reflected spectrum is a composite of emission from the receding red and approaching blue sides which leads to separation in line energies as a result of Doppler shifting and a classic `double-horned' profile; as these radii approach the BH, the radial velocity increases which leads to a greater separation. As the orbiting material near the BH is mildly relativistic, beaming of the emission (resulting from relativistic aberration and time dilation) leads to a flux change of a factor D$^{3}$ where D is the Doppler factor:

\begin{figure}
\hspace{0.5cm}
\includegraphics[scale=.5]{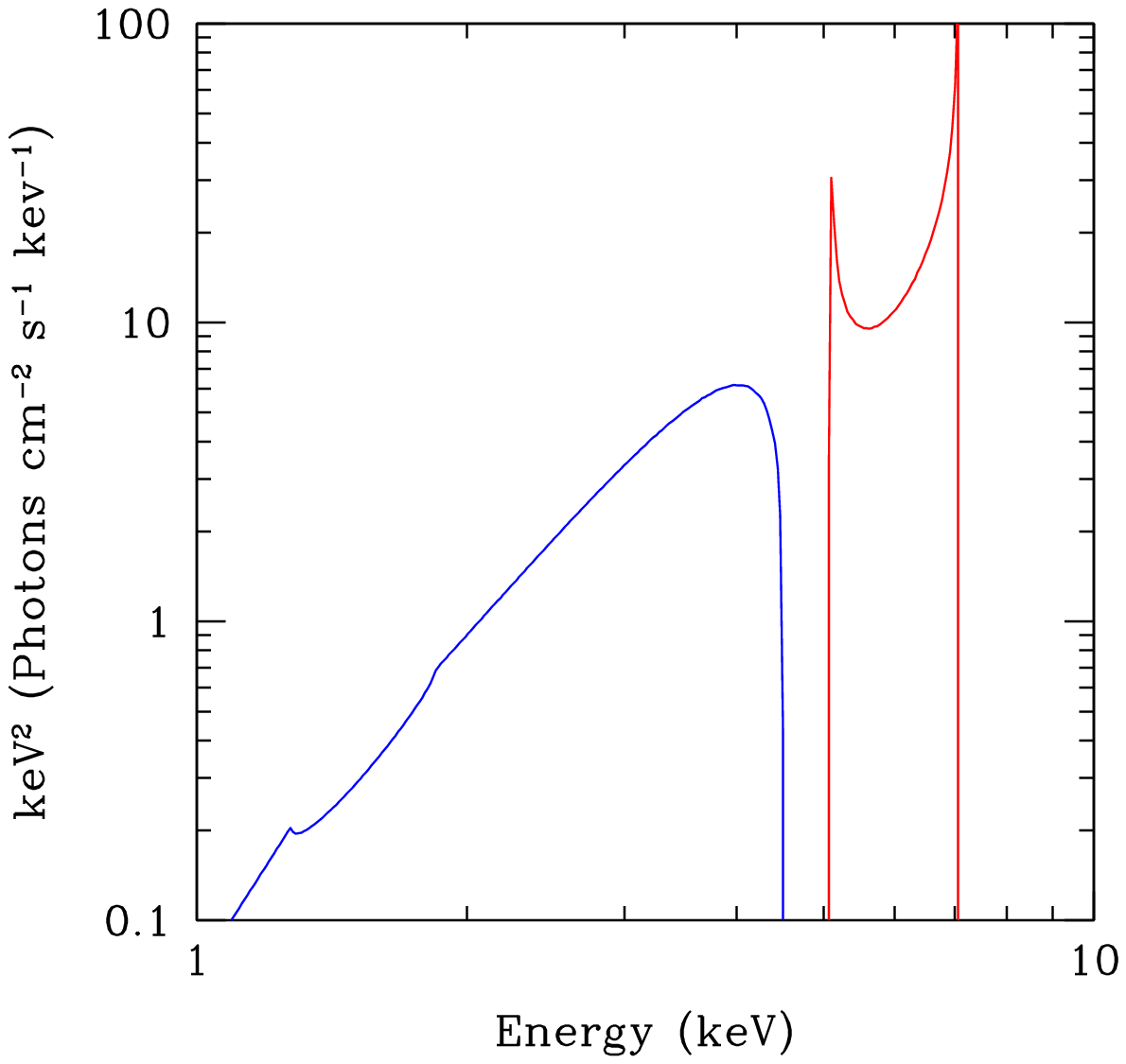}
\caption{Distortion of an emission line (using the model {\sc relline}: Dauser et al. 2010) for two annuli in the accretion disc seen at 45 degrees about a BH with a$_{*}$ = 0.998. The red line is from an annulus 20-22 R$_{g}$ from the BH and is broadened into the characteristic `double-horned' profile by Doppler shifting. The blue and red `wings' of this line are then increased and decreased in flux respectively through the effect of Doppler boosting (relativistic beaming: see equation 34). The blue line comes from an annulus at the ISCO (1.3-2 R$_{g}$ from the BH) where the red wing gives a measure of the position of the ISCO; the entire line is shifted to lower energies due to the combination of transverse Doppler shift and gravitational redshift. In reality we see a blend of lines from all radii.}
\end{figure}

\begin{equation}
D = \{\Gamma[1-\beta cos(i)]\}^{-1}
\end{equation}
$\Gamma$ is the Lorentz factor, $\beta = v/c$, $v$ is the approaching or receding velocity and $i$ is the inclination of the observer from the rotation axis. As a result of relativistic beaming, flux from the approaching, blue-shifted side is Doppler boosted whilst the receding side is Doppler de-boosted (see e.g. Fabian et al. 1989, 2000; Stella 1990). Accompanying these Newtonian and special relativistic effects are the special relativistic effect of transverse Doppler shift and the general relativistic effect of gravitational redshift, both of which act to reduce the observed energy of the line at each radius. It can be seen that the combination of these effects leads to a heavily skewed and broadened line, where the blue wing depends heavily on the inclination (equation 34) and the shape of the red wing is dominated by the position of the inner edge and can therefore be used as a proxy for the BH spin (e.g. Laor 1991). In Figure 3 we demonstrate these effects on a fluorescence line originating from two annuli in the disc. Importantly, unlike the case with continuum fitting, the method of fitting the relativistically broadened Fe line is {\it independent} of BH mass. In addition, the technique is insensitive to the distance to the source and the inclination is a ubiquitous parameter of the spectral models (determined from the shape of the blue line and Fe K$_{\alpha}$ absorption edge) and so can be estimated concurrently with the the spin. As a consequence, the major source of error when determining the spin is the systematic uncertainty within the model itself. Of importance when determining the emergent reflection spectrum is the modelling of the primary illuminating spectrum which may be more complex than a simple power-law but can be probed through the use of an expanded bandpass (e.g. the {\it NuSTAR} observations of Cyg X-1: Parker et al. 2015) as well as through application of advanced spectral-timing techniques (see the view of Uttley et al. 2014). Additionally, the radial and vertical dependent density structure of the optically thick disc, and the geometry of the corona/disc system (which sets the emissivity e.g. Ghisellini, Haardt \& Matt 1994) are important considerations for accurate modelling of the reflected emission. The following subsections discuss these issues in turn.

\subsubsection{Emissivity \& Geometry:}
An important ingredient of the reflection spectrum is the `emissivity' ($\epsilon$) of the flux from the disc which is proportional to the radial dependence of the illuminating radiation {\it onto} the disc (where the irradiation goes as $I(R) \propto R^{-\epsilon}$). For an irradiating point source in flat space-time, sitting above the disc, the emissivity goes as the product of the inverse square law and the cosine of the normal to the plane of the disc (see the discussion in Wilkins \& Fabian 2012) or, for a source, $h$ above the disc, $\epsilon$ goes as  $(R^{2}+h^{2})^{-1}*cos\theta$ (where $cos\theta$ is just $h/\sqrt{(R^{2} + h^{2})}$). Thus at small radii ($R \ll h$) this would tend to a flat profile but at large radii (around $R = h$) will instead tend to $R^{-3}$ which is that of the disc emission in equation 27. This discounts the effect of relativity however, and the effect of light-bending close to the BH can have a significant impact by focussing more of the radiation onto smaller radii. 

\begin{figure}[h]
\sidecaption
\hspace{2cm}
\includegraphics[scale=.75]{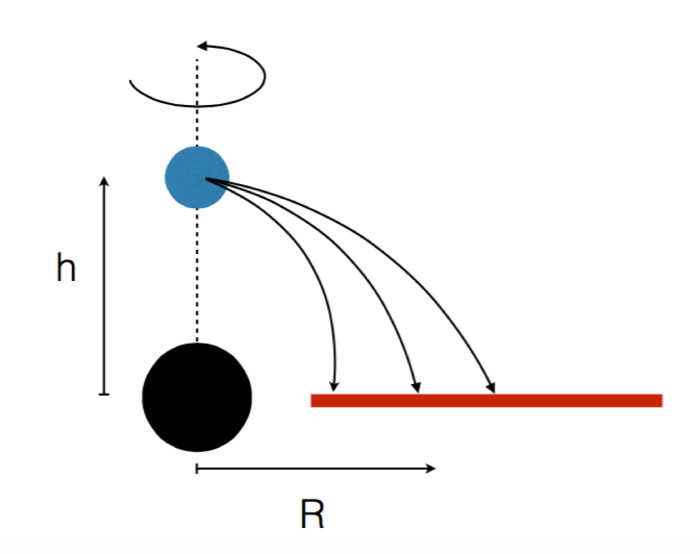}
\caption{Lamp-post geometry showing the X-ray source (blue) at some height $h$ above the BH (on the spin axis) as expected for the base of a jet. The X-ray photon paths are bent by the strong gravity and illuminate the disc, leading to a steeper emissivity profile at small radii than a simple flat profile (which breaks to $R^{-3}$ at large disc radii) otherwise expected from illumination in flat space-time.}
\label{fig:2}       
\end{figure}

The impact of light-bending naturally depends on the location of the corona and the geometry of the disc-corona system. One of the most popular geometries is that of the `lamp-post' (see Figure 4) where the corona sits on the BH spin axis some height above the BH; in a physical sense this would then be associated with the base of a jet (see Markoff \& Nowak 2004; Dauser et al. 2013; Wilkins \& Fabian 2012; Wilkins \& Gallo 2015). The result of light-bending in this geometry can lead to highly anisotropic illuminations, and is predicted to produce a radial profile that is a twice-broken power-law, with very steep emissivities in the most inner regions then flattening before tending to constant emissivity at large radii (Miniutti et al. 2003; Martocchia, Karas \& Matt 2000). Naturally, as the profile is a function of the light-bending, it is a function of the source height above the disc. For a decrease in height, the emissivity profile is steepened due to an increased amount of light-bending, i.e. a larger value of $\epsilon$ (Reynolds \& Begelman 1997; Miniutti \& Fabian 2004; Wilkins \& Fabian 2012), which drops the observed flux from the corona and increases the amount of reflection, the ratio of which is the `reflected fraction'. In addition, the position of the break to a flatter emissivity law changes as does the extent of the region over which this is predicted to hold (Wilkins \& Fabian 2012).  Although reflection models without ray-tracing (or incorporating a relativistic transfer integration kernel) cannot account for this effect {\it directly}, convolution models which take into account the effect of photon orbits in the Kerr metric have been developed and can be used to alter the emissivity profile accordingly. The most commonly used models (besides {\sc relconv} which we will mention shortly in the context of next-generation models) are {\sc kdblur/kdblur2} which are based on the relativistic transfer function of Laor (1991) and {\sc kerrconv} which uses an analytical prescription for relativistic beaming and a pared down transfer function for ease of computing (Brenneman \& Reynolds 2006) 

The geometry of the illuminating source remains a fundamental issue in accretion physics in general but has especially  important relevance for reflection models (although as we will discuss there are time-domain methods now in use to help constrain this) as this sets the possible emissivity profiles. It
may well be lamp-post-like in the picture where the optically-thin
electron population is associated with the base of a jet (and, if the
component is analogous to that in BHBs, probably composed of a thermal electron
population). In the case where the scattering is not in a jet base,
other geometries must be considered. Wilkins \& Fabian (2012) explored a series of potential geometries for the corona, notably testing not only for position of the illuminating source but also for the source's extent which had not been previously considered from a theoretical perspective. They calculated emissivity profiles from ray-tracing using GPUs (graphics processing units) and include a full treatment of the relativistic effects including the reduction in effective disc area (which they find goes as the classical disc area divided by the redshift: see their Figure 2) and blue shifting of photons onto the disc. The geometries considered include an axial source (lamp-post geometry) both stationary and moving away from the BH (i.e. a jet), an orbiting source (which can be seen as a single element in a ring) and an extended disc-like source. In the case of the ring source, the act of moving it further from its position above the BH leads to a somewhat reduced emissivity profile in the central regions with a flatter profile after the first break and extending to larger radii, dependent on the radial position of the ring. As there are several arguments for an extended corona from both variability studies (e.g. Churazov, Gilfanov \& Revnivtsev 2001) and more directly from gravitational micro-lensing (which indicates an extension of up to a few R$_{\rm g}$: Dai et al. 2010), an isotropic point-source is unlikely to be representative and to this end, the study of an extended source is extremely valuable. Wilkins \& Fabian (2012), describe such a scenario as the sum of points spread over a vertical (giving a source of finite thickness) and radial extent (i.e. a disc above the disc). The source is assumed to have a constant luminosity across its extent and is optically thin to the radiation it has produced (and so does not interact with itself); as expected, the resulting emissivity profile is a combination of the breaks and indices from consideration of point sources (although may well be further complicated if magnetic reconnection occurs at distinct radii in the disc, e.g. Sochora et al. 2011). Wilkins \& Fabian (2012) explore the effect of changing the spin on the emissivity
, finding that the increased disc area resulting from a higher spin leads to a steeper emissivity for an axial or ring geometry (with Wilkins \& Fabian 2012 showing that the emissivity is only greatly steepened in these arrangements for $a_{*} \ge$ 0.8). 

Finally, in addition to the issues associated with the as-yet-unknown coronal geometry, a further simplifying assumption commonly made is that the coronal plasma itself is uniform, being of a single temperature and density. This is unlikely to be the case (see e.g. Parker et al. 2015) and in future can be tested using eclipses to determine the radial dependence of the coronal properties (see section 5).

%

\subsubsection{Disc vertical density profile:}

As with estimating the BH spin via the disc emission (see previous section), the reflection spectrum also depends on the vertical structure of the disc; in the former, the important consideration was the position of the effective optical depth (which, as we discussed, can change depending on the relative dominance of opacities). In the case of reflection spectra, the important consideration is the ionisation state of the material in which the majority of the reflection takes place (i.e. one effective optical depth). Shakura \& Sunyaev (1973) point out that in radiation pressure dominated discs of large optical depth and heated by viscous dissipation, the density in the vertical direction is roughly constant; this has led to models for both AGN (Ross \& Fabian 1993; Matt, Fabian \& Ross 1993; Zycki et al. 1994) and BHBs (Ross, Fabian \& Brandt 1996; Ross, Fabian \& Young 1999) incorporating this simplifying assumption of a constant vertical density structure in the disc (or a Gaussian distribution: Ross, Fabian \& Young 1999). This assumption has been questioned by Nayakshin, Kazanas \& Kallman (2000) who point out that the illuminating X-ray heating of the outer disc layers (where reflection occurs) could be orders of magnitude greater than the viscous heating in the disc at such height above the mid-plane. The disc material is expected to be thermally unstable when the supporting gas pressure in the illuminated atmosphere falls below a fraction of the downwards radiation pressure provided by the illuminating continuum (Krolik, McKee \& Tarter 1981; Raymond 1993; Ko \& Kallman 1994; R{\'o}{\.z}a{\'n}ska \& Czerny 1996; R{\'o}{\.z}a{\'n}ska 1999) which leads to a large jump in temperature. Nayakshin, Kazanas \& Kallman (2000) argue that a self-consistent determination of the density gradient is therefore necessary and determine the density profile from the condition of hydrostatic equilibrium, simultaneously solving the equations of ionisation, energy balance and radiative transfer, finding constant density (or Gaussian distributions) to be broadly unphysical. Their calculations instead suggest that the structure of the illuminated layer of the disc is a two phase structure formed of an ionised skin where, in the case of `hard' illuminating flux, Fe is completely ionised and below this layer the material is cold, i.e. weakly ionised. The strength of the reflection features is then a function of the optical depth of the top layer (which does not imprint features due to being totally ionised), with smaller optical depths of the skin leading to less of an impact on the emergent reflection spectrum from the cold layers beneath (as such layers are reached more readily). This in turn depends on the strength of the illumination and is therefore intrinsically connected to the emissivity profile discussed above. As pointed out by Fabian \& Ross (2010), simulations of discs supported by magnetic pressure (e.g. Blaes, Hirose \& Krolik 2007) add a complicating factor, as this can potentially reduce the density and increase the impact of photo-ionisation. Indeed, should magnetic pressure dominate the hydrostatic support then the discontinuous vertical profile resulting from the thermal instability may not exist, although large scale inhomogeneities could be present and these could have an impact on the emergent reflection spectrum (Ballantyne, Turner \& Blaes 2004; Blaes, Hirose \& Krolik 2007). As the impact of magnetic fields on the disc structure remains an open question, so too does the nature of hydrostatic balance.

\subsubsection{Effect of the plunging region:}
 
A standard assumption in modelling the emission from the BH accretion flow is that truncation at the ISCO is final with no radiation arriving to the observer from the ballistically infalling material in the plunging region. Zhu et al. (2012) studied this low density region through 3D MHD simulations, discovering that only a small amount of thermal emission was produced and was unlikely to distort spin values obtained via the continuum fitting method (section 3.1). 

In order to determine the effect of including the plunging region on the {\it reflected} spectrum (as first discussed in Reynolds \& Begelman 1997), Reynolds \& Fabian (2008) performed high resolution 3D MHD simulations (thereby incorporating MRI driven turbulence: Balbus \& Hawley 1991; 1998) of the geometrically thin accretion disc close to the ISCO. Their simulation demonstrates that some Fe K$_{\alpha}$ emission can originate  from the plunging region which introduces an additional systematic error for models which terminate at the ISCO. As a notable caveat to this effect, at high mass accretion rates the density in this region drops and the material may become ionised to the point where there are no longer any notable reflection features (with the exception of the Compton hump). The authors estimate the uncertainty introduced by the inclusion of reflection from within the plunging region (see their Figure 5), finding it to be an overestimate of the true spin (as for the inclusion of emission from this region in continuum fitting, e.g. Zhu et al. 2012) with a larger error for slowly spinning BHs, and to be relatively insensitive to uncertainty in the inclination (as long as the inclination can be constrained from modelling the Fe K$_{\alpha}$ profile). The uncertainty grows when the scale height is large just outside of the ISCO; whilst further work is needed to establish the full impact of this, Reynolds (2014) notes that inclusion of such uncertainties can relax the constraints for otherwise maximal spinning BHs to $>$ 0.9.


\subsubsection{Next-generation models:}

The most recent codes (developed principally by Thomas Dauser, Javier
Garc{\'{\i}}a and colleagues) provide important updates to existing
models. In the case of non-relativistic reflection (i.e. no light
bending effects) {\sc xillver} (Garc{\'{\i}}a \& Kallman 2010; Garc{\'{\i}}a et
al. 2011; Garc{\'{\i}}a et al. 2013) includes {\sc xstar} (Kallman \&
Bautista 2001) to solve for the ionisation state of the disc
atmosphere and makes use of the most updated, accurate and complete
atomic database for atomic transitions. A series of
codes have also been developed which include a relativistic transfer
function from ray-tracing which allows the
emissivity of the irradiation to be determined (although the disc is
still assumed to be `thin' as in the Novikov-Thorne
prescription). These models include line emission for spins ranging
from maximal retrograde through to maximal prograde ({\sc relline}:
Dauser et al. 2010) and incorporate limb-darkening, limb brightening
and is similar in output to previous models including {\sc kerrdisk}
and {\sc kyrline} (Dov{\v c}iak, Karas \& Yaqoob 2004) but is evaluated over a finer energy grid than {\sc
  laor}. This model has been adapted to act as a convolution kernel
({\sc relconv}) for the entire spectrum which, when combined with {\sc
  xillver} allows the entire reflection spectrum to be calculated with
a prescribed emissivity law ({\sc relxill}: Dauser et
al. 2013). Using this last model, Dauser et al. (2014) showed that, by
considering the additional spin-dependence of the reflected fraction
(which most models do not account for), it becomes possible to place
increasingly stringent constraints on the spin by discounting unphysical
solutions (see also Parker et al. 2014), although consideration of distant, neutral reflection may
complicate matters somewhat. In addition to improvements in determining the emissivity, {\sc relxill} also allows the thermal cut-off (at energies often well beyond the detector bandpass) to be determined (Garcia et al. 2015) whilst calculating the angular dependence of the reflection spectrum (approximated in the past by a convolution of the angle-averaged reflection spectrum with a relativistic kernel) in a fully self-consistent manner. The latter has only a minor impact on measurements of the spin and inner disc inclination (although the constraints on both parameters improves) but can have a major impact on the estimated Fe abundance (Garcia et al. 2014). Although an undoubted step forward, this
model still relies on several caveats that are important to consider;
the radial and vertical density profile is assumed constant, as is the ionisation state of
the reflecting material. 
Further iterations of the {\sc relxill}\footnotemark\footnotetext{www.sternwarte.uni-erlangen.de/$\sim$dauser/research/relxill/} model have begun to
deal with these assumptions, by considering either a power-law-like
radial density structure, multi-zone ionisation structure or with the
ionisation gradient calculated self-consistently from irradiation.


\subsubsection{The geometry through reverberation:}   

As the geometry of the disc-corona system is potentially important for reliably estimating the spin, we will very briefly discuss the role of reverberation as the most promising method for placing tight (spectrally independent) constraints. 

It was realised early on (Fabian et al. 1989; Stella 1990) that the physical separation of the corona and the accretion disc will, from the perspective of an observer at infinity, lead to a light-travel time delay between the emission arriving directly from the corona and that reflected from the disc (see Matt \& Perola 1992; Campana \& Stella 1993; Fabian et al. 2000); a simple schematic showing this for the lamp-post geometry is given in Figure 4. This `reverberation' off the disc can be studied in the time domain from the cross-correlation function (e.g. Gandhi et al. 2010) between a band containing the intrinsic hard emission (from the corona) and a band dominated by reflection. For ease of evaluating the lag across multiple frequencies, reverberation is more commonly studied in the Fourier domain via `phase lags'. As the phase lag is a function of the geometry, it can provide important insights into the nature of the accretion flow and has implications for how we measure the spin. In practice, obtaining the phase lag requires evaluating the components of the cross spectrum (Vaughan \& Nowak 1997; Nowak et al. 1999) which in turn are found from the frequency-dependent complex ordinates of the Fourier transformed lightcurve. Here we present only the most basic of descriptions and for a detailed review of Fourier analysis we point the reader to van der Klis (1988) and for a detailed review of reverberation and phase lags we recommend that of Uttley et al. (2014).

The lightcurve in each energy band (e.g. those dominated by either the primary and reflected emission) is given by $x(t)$ and $y(t)$ with their Fourier transforms $X(\nu)$ and $Y(\nu)$. An alternative way of displaying these is $X(\nu) =  A(\nu)e^{\! i\Psi}$ and $Y(\nu) =  B(\nu)e^{\! i\Psi + \phi}$ where $\phi$ is the phase lag between them. By defining the cross-spectrum as $C(\nu) = X^{*}(\nu)Y(\nu)$ (Nowak \& Vaughan 1997) the phase can then be determined from:

\begin{equation}
\phi(\nu) = {\rm arg}[C(\nu)] = {\rm tan}^{-1} \frac{\Im(\nu)}{\Re(\nu)}
\end{equation}
where  $\Re$ and $\Im$ are the real and imaginary Fourier coefficients of the cross spectrum at each frequency. The details of how to estimate the phase in practice (i.e. taking averages over segments and normalising), are discussed in Uttley et al. (2014). The important question is how do we go from the phase as measured from the lightcurve to the geometry? Each lightcurve can be related via an impulse response to an input lightcurve, called a `driving signal', $s(t)$, such that:
 
\begin{equation}
x(t) = \int_{-\infty}^{\infty} \! h(t-\tau)s(\tau) {\rm d}\tau
\end{equation}
where $\tau$ is the time lag ($\phi/2\pi\nu$). In Fourier space this is equivilent to saying  $X(\nu)$ = $H(\nu)S(\nu)$ (and similarly, $Y(\nu)$ = $G(\nu)S(\nu)$), where $H(\nu)$ (or equally $G(\nu)$) is the Fourier transform of $h(t-\tau)$ and is called the transfer function. As described in Uttley et al. (2014), the impulse response (and therefore the geometry) is related to the phase lag in equation 35 via the cross spectrum:

\begin{equation}
C(\nu) =  H^{*}(\nu)S^{*}(\nu)G(\nu)S(\nu) = |S(\nu)|^{2}H^{*}(\nu)G(\nu)
\end{equation}

Thus the cross-spectrum of two lightcurves contains the cross spectrum of the transfer functions ($H^{*}(\nu)G(\nu)$) and the power spectrum of the driving signal. As the power spectrum of the driving signal ($|S(\nu)|^{2}$) is by definition  real valued it has no effect on the phase and so, given an input relativistic transfer function from the geometries described in section 3.2.1, it therefore becomes possible to compare the expected phase lag to observations and thereby begin to constrain potential geometries (e.g. Cackett et al. 2014). In so doing, such analyses also hold the promise of better understanding contributory factors in measuring the spin. 


As reverberation in BHBs occurs on very fast timescales, the number of photon counts per light crossing time is small; as such, observations of reverberation due to reflection have to-date focussed mainly on AGN (where conversely, the number of photon counts per light crossing time is substantial). The first tentative hints of a signal were noticed by McHardy et al. (2007) with the first significant ($\ge$ 5-$\sigma$) detection of a reverberation lag due to reflection made by Fabian et al. (2009) from {\it XMM-Newton} observations of the AGN, 1H0707 (see also Zoghbi et al. 2010). Due to the remarkably strong Fe L emission in this source (which is noted for having super-solar abundance of Fe), the authors were able to take a hard band without strong contributions from reflection and detect the signature of reverberation as a `soft lag', i.e. the soft emission lagging the hard. Notably the strong Fe K$_{\alpha}$ emission seen in the energy-lag spectra (e.g. Zoghbi et al. 2012; 2013; Kara et al. 2013a; 2014; 2015a) and its continuation to higher energies mapping out the Compton hump (through use of {\it NuSTAR} data: Zoghbi et al. 2014; Kara et al. 2015b) on the frequencies of the reverberation lag all indicate that its origin lies in relativistic reflection.


Following the initial discovery of reverberation, a soft lag has now been detected (or hinted at) in several AGN (e.g. Emmanoulopoulos, McHardy \& Papadakis 2011) with the detection of a significant trend of frequency/amplitude with mass (De Marco et al. 2013). This would require that the coronal geometry be the same throughout, although in at least one AGN (IRAS 13224-3809: Kara et al. 2013b) the soft lag is observed to change in frequency and amplitude; this does not necessarily invalidate a correlation but understanding the stability of the corona-disc geometry is clearly of great importance.

Work is now turning towards reconstructing the spin-dependent geometry (see section 3.2.1 and Wilkins \& Gallo 2015 for a recent study of this) through use of impulse response functions (e.g. Reynolds et al. 1999) and {\it direct} modelling of the phase lags. Young \& Reynolds (2000) and more recently Cackett et al. (2014) and Emmanoulopoulos et al. (2014) simulate the effect of reverberation assuming a lamp-post geometry and obtain the associated impulse response functions for a variety of physical parameters (either via pre-calculated transfer functions in the case of the former or via direct ray-tracing in the case of the latter). The major effect on the frequency-dependent lag results from changing the mass or vertical displacement of the axial illuminator and is relatively insensitive to the inclination and spin. The energy profile of the lag however is far more sensitive, with the profile of Fe K$_{\alpha}$ - found to be strong at frequencies corresponding to the soft lag - changing with spin in a similar manner to its time-averaged counterpart. Cackett et al. (2014) compare predicted frequency-lag and energy-lag spectra to the real data of NGC 4151 and constrain the physical parameters including the spin (found to be maximal), inclination and source height. In a similar approach, Emmanoulopoulos et al. (2014) perform the first sample analysis, modelling the frequency-lag spectra in 12 AGN using general relativistic impulse response functions, finding a consistent source height across the sample and a possible bi-modality of spin values (which might indicate SMBH growth via mergers, e.g. Volonteri et al. 2013). In order to reliably estimate the spin, these techniques typically rely on high quality data with high energy resolution; whilst not widely available at present, the arrival of next-generation missions such as ESA's {\it Athena} will allow such methods to be used to their full potential. In addition, Athena's location at the L2 point will allow uninterrupted observations which will allow lower frequencies to be explored, important for larger mass AGN where the reverberation signal is expected to be found (e.g. De Marco et al. 2013).



\subsection{Results: BHBs}

Here we will discuss the results of campaigns to estimate the spin of stellar mass BHs from applying the methods described in sections 3.1 and 3.2. Due to their success, the number of sources for which spin measurements have been obtained is constantly growing and we apologise for any results which are therefore absent. Whilst we briefly touch upon the relevance of measuring the BH spin for the purposes of probing the launching of astrophysical jets, obtaining an understanding of the spin distribution is important in it's own right as this provides a view of the natal spin set during the supernova process (as insufficient mass can been accreted onto the BH to change the spin by a considerable amount: see the discussion of Reynolds 2014 and section 2). 

\subsubsection {Continuum fitting}

As discussed elsewhere in this compilation, the outburst of BHBs (with low mass companion stars), follows a predictable path through X-ray spectra, variability and multi-wavelength properties (Fender, Belloni \& Gallo 2004; Fender, Homan \& Belloni 2009). The luminosity of a source is usually parameterised as a ratio to the Eddington limit for spherical accretion, given by:

\begin{equation}
 L_{\rm Edd} = \frac{4\pi GMm_{\rm p}}{\sigma_{T}} 
\end{equation}
where $\sigma_{\rm T}$ is the Thompson cross section and $m_{\rm p}$ is the proton mass. This is frequently referred to by its numerical approximation (and under the assumption that the material is entirely ionised Hydrogen), L$_{\rm Edd}$ = 1.26$\times$10$^{38}$ M$_{\rm BH}$/M$_{\odot}$. 

On the rise to outburst, at low to moderate mass accretion rates (typically $<$ 70\% of the Eddington limit: Dunn  et al. 2010), the spectrum is dominated by a hard tail of emission resulting from inverse Compton scattering by an optically thin, thermal population of electrons (thus the spectrum is referred to as being in a hard state). This emission is accompanied by synchrotron emission extending across several decades in frequency and is associated with a low bulk-Lorentz factor ($\Gamma$) jet. At the highest mass accretion rates, the spectrum begins to soften and becomes increasingly dominated by emission from the accretion disc, passing through the intermediate states and then into the soft state (we note that there are accompanying changes in the variability properties but these are not of relevance for the discussion here). 

The soft state (also called the thermal dominant state) is characterised by a strong disc component and a non-thermal (rather than thermal) tail of emission to high energies which is only of the order of a few percent or less of the total flux in the X-ray bandpass (e.g. McClintock \& Remillard  2006; Remillard \& McClintock  2006). As the vast majority of the emission originates from the disc (with very little energy being liberated in a corona), the spin can in principle be determined from the application of a suitable model for the disc emission (see section 3.1). However, a condition of applying this technique relies on the inner edge being located at the ISCO (as expected from GRMHD simulations: Shafee et al. 2008a; Penna et al. 2010). As the luminosity of the disc component for a fixed emitting area (i.e. fixed inner edge) is expected to follow a T$^{4}$ dependence (see equation 27), this provides a means by which to test the consistency of the position of the ISCO. Determining the position and stability of the ISCO via this approach has been attempted by a number of authors (e.g. Ebisawa et al. 1991; Muno, Morgan \& Remillard 1999; Kubota \& Done 2004; Kubota \& Makishima 2004; Steiner et al. 2010). In particular, Gierli{\'n}ski  \& Done (2004) use the expected relation for the integrated disc luminosity as a function of the maximum observed colour temperature (Gierli{\'n}ski  et al. 1999) considering a pseudo-Newtonian potential (Paczynski \& Wiita 1980) and corrections to the observed flux due to inclination and GR effects (Cunningham 1975; Zhang, Cui \& Chen 1997) to obtain the predicted form of the relation between Eddington ratio (L/L$_{\rm Edd}$) and temperature. Through use of RXTE data (which covers a nominal 3-20~keV energy range) for a number of well known BHBs, Gierli{\'n}ski \& Done (2004) showed that the disc emission in the soft state is consistent with a fixed inner edge for a range of Eddington ratios, with deviations at the very highest and lowest values. As the predicted relation depends on GR corrections and therefore the spin, Gierli{\'n}ski \& Done (2004) were able to use these plots to indicate that the spin in the case of XTE J1550-564 is non-maximal.

The first attempts to {\it directly} constrain the spin from modelling the disc emission (and indeed pioneering the field of continuum fitting) were carried out by Zhang et al. (1997). The authors utilised values for the peak temperature and flux for a number of BHBs available from the literature (e.g. Dotani et al. 1997; Belloni et al. 1997) to obtain the position of the inner edge after accounting for relativistic effects. Following the development of new models (namely {\sc kerrbb} and {\sc bhspec}) there has been a steady and substantial increase in spin measurements obtained via this method, with 12 in total covering persistent and transient sources both within our Galaxy and in nearby galaxies. McClintock, Narayan \& Steiner (2014) provide a review of the continuum fitting method and its application up to 2013 and we highly recommend this as an excellent overview.

As mentioned in section 3.1.1, of critical importance in all attempts to constrain the spin via the continuum fitting method has been the accuracy with which the system parameters - the inclination, distance and BH mass - can be determined as the uncertainty on these dominates over typical model systematics of $\sim$5\%. As noted by Orosz et al. (2011); having an accurate distance to the system is key to reducing uncertainty on the mass estimate. For those systems in nearby galaxies, cosmic distance ladders can be used and result in $\sim$a few percent uncertainty, substantially better than typical accuracies of distance measurements for Galactic sources (Jonker \& Nelemans 2004), although these are being substantially improved through radio parallax measurements (see Miller-Jones 2014). Once the distance is known, the mass can be estimated from modelling the orbital dynamics of the system using eclipsing light curve (ELC) models (e.g. Orosz \& Hauschildt 2000, Orosz et al. 2011) and requires the radial velocity of the companion star derived from line spectroscopy. Such modelling also determines the inclination of the system, however, the question remains as to whether the system inclination is representative of the inclination of the inner disc which is key for determining the spin via the continuum fitting method. As discussed in section 2.1.1, misalignment of the BH spin axis with that of the binary orbit can lead to relativistic precession and the Bardeen Peterson effect which aligns the inner regions with the BH spin axis, whilst the outer regions align with the binary plane. There may already be evidence in support of this scenario; the inclination of the jet, which is expected to be the same as the BH spin axis, appears misaligned with the binary orbit in the BHBs GRO J1655-40 (Hjellming \& Rupen 1995; Greene, Bailyn \& Orosz 2001 and also Mirabel et al. 2002 for possible issues associated with accurately determining such physical properties) and SAX J 1819-2525 (also known as V4641 Sgr: Maccarone 2002, Martin, Reis \& Pringle 2008). As an interesting aside, it is possible that the ubiquitous low frequency quasi-periodic oscillations (LFQPOs) seen in the variability power spectra of BHBs when in the hard through to the intermediate states (see the review of Belloni \& Stella 2014), are associated with precession of a low density flow in the inner regions due to misalignment (Ingram, Done \& Fragile 2009; Ingram \& Done 2011). Incorporating the effect of reflection of the radiation from the precessing regions leads to unambiguous, observational tests for misalignment which, if confirmed, would have implications for our ability to measure the spin reliably (see Ingram \& Done 2012). We will return to the observational effect of precession in section 4.

In obtaining the spin via the continuum method, a further selection criteria is often applied to the X-ray spectra. As discussed in McClintock et al. (2006), above $\sim$30\% of the Eddington limit, the spectra may deviate from that expected from a simple thin disc possibly due to the creation of an inner, optically thick, radiation pressure dominated corona/slim disc due to the high mass accretion rates (Abramowicz et al. 1988; Ueda et al. 2009; Middleton et al. 2012; 2013) or a region supported by magnetic pressure (Straub, Done \& Middleton 2013). It is assumed that either disc truncation or simply a cooling of the disc photons by the corona leads to a lower disc temperature than should be expected from the innermost edge of the disc and a deviation away from the expected $L \propto T^{4}$ relation (Gierli{\'n}ski  \& Done 2004). Such deviations in disc structure can in principle lead to the spin being underestimated and has been proposed to explain the difference in spin results for the extreme BHB, GRS 1915+105 with both moderate (Middleton et al. 2006) and maximal (McClintock et al. 2006) values claimed. However, the actual luminosity at which this distortion appears in GRS 1915+105 is model dependent and requires careful treatment (Middleton et al. 2006); as a result, the spin of this unusual source remains somewhat contentious. In general, the point at which radiation pressure starts to affect the structure of the flow must be related to the spin (which sets the radiative efficiency - see equation 30) and so the limit of 30\% of the Eddington limit translates into differing mass accretion rates for different sources. Should the structure of the disc be affected instead by something coupled to the mass accretion rate, the change in the flow could conceivably occur at a different Eddington ratio; thus to ensure rigour, selection criteria should ideally be determined on an object-by-object basis (e.g. Steiner et al. 2010). 

An important consideration for spectral studies of BHBs (in general but in particular in using the spectrum to estimate the spin) is that, whilst their proximity leads to high fluxes, these can lead to severe issues for CCD spectrometers (see Kohlemainen, Done \& D{\'{\i}}az Trigo 2014 for a discussion of these effects on {\it XMM-Newton} data) including photon pile-up (where the arrival of multiple photons is read as a single event), charge transfer inefficiency (CTI, where a loss of charge occurs during CCD readout) and X-ray loading (where very bright sources contaminate the ``offset map" - analogous to a ``dark frame" in optical instruments). Such effects can readily distort the spectrum and therefore care must be taken to ensure that their impact on estimating the spin is understood.

Finally, it is important to note that mass loaded winds are launched from the accretion disc (perhaps as a result of radiation pressure, e.g. Proga \& Kallman 2002; thermal reprocessing in the outer disc, e.g. Begelman, McKee \& Shields 1983 or MHD driving, e.g. Neilsen \& Homan 2012) and become stronger as the source becomes dominated by thermal disc emission (Ponti et al. 2012). Although these winds are highly ionised (as the source is X-ray bright) and are unlikely to present an obstacle for studying the continuum (although see the following section for their impact on AGN spectra), the models applied in order to determine the spin (see section 3.1.1) are at present only steady-state and do not take this mass loss into account - the effect of mass loss on the measurement of the spin at this time is therefore unknown. 

In Table 1 we present the BH spin values for the present sample of BHBs (both Galactic and extra-galactic) along with the system parameters and the model used (useful in light of the assumptions that underpin each model as discussed above). The vast majority of these values are reported in the recent review by Miller \& Miller (2015) and for the sake of brevity we direct the interested reader there for the individual references for the spin values and system parameters, although we note where possible, updated values and any additional sources. 

Notably, all bar two of the BHBs in Table 1 are Galactic (or located in the large Magellanic cloud) and as such the issue of precision in the distance measurement is important. In the case of the source in M31 (Middleton et al. 2013; 2014), the distance is known to within a few percent and as this is relatively large, the source flux (which is high in such soft states) does not pose an issue for CCD detectors. Thus such bright (yet relatively nearby) extra-galactic sources offer a means to expand our sample of BHBs with spin measurements which, as we discuss in section 3.5 may be extremely important for understanding how astrophysical jets are launched.

\begin{table*}
\begin{center}
\begin{minipage}{120mm}
\caption{Table of BHB spins}
\begin{tabular}{lclclclclc|c|c|}
\hline
\hline
& \multicolumn{5}{c}{Continuum fitting} & Reflection fitting\\
\hline
Source & Mass & Inclination & Distance & \hspace{0.5cm} a$_{*}$ & Model & \hspace{0.5cm} a$_{*}$  \\ 
	     & (M$_{\odot}$)  & (degrees) &  (kpc) &  &  {\sc K/K2/B}	      & 	\\
	     \hline
Cygnus X-1 & 14.8 $\pm$1.0 & 27.1 $\pm$ 0.8 & 1.86 $^{+0.12}_{-0.11}$ &$\ge$ 0.95 & K2 & $>$ 0.97 $^{a}$\\
XTE J1550-564 & 9.10 $\pm$ 0.61  & 74.7 $\pm$ 3.8 &  4.38$^{+0.58}_{-0.31}$ & 0.34 $^{+0.37}_{-0.34}$ & K2  & 0.55 $\pm$ 0.22	\\
XTE J1650-500 &  &  &   & &  	      & 0.79 $\pm$ 0.01	\\
XTE J1652-453  &   &  &   &  &  	      & 0.45 $\pm$ 0.02	\\ 
XTE J1752-223 &  & &  &  &  	      & 0.52 $\pm$ 0.11	\\
XTE J1908+094 &  &  &  &  &  	      & 0.75 $\pm$ 0.09	\\
A 0620-00 & 6.61 $\pm$ 0.25  &  51.0 $\pm$ 0.9 &  1.06 $\pm$ 0.12 & 0.12 $\pm$ 0.19 &  	K2      & 	\\
4U 1543-475 & 9.4 $\pm$ 1.0  & 20.7 $\pm$ 1.5 &  7.5 $\pm$ 1.0 & 0.8 $\pm$ 0.1 & K 	      & 0.3 $\pm$ 0.1	\\
4U 1630-472 &  &  &  &  &  	      & 0.985 $^{+0.005}_{-0.014}$	\\
MAXI J1836-194 &   &  &  & &  	      & 0.88 $\pm$ 0.03	\\
GRO J1655-40 & 6.30 $\pm$ 0.27  & 70.2 $\pm$ 1.2 & 3.2 $\pm$ 0.2 & 0.7 $\pm$ 0.1 &  K & 0.98 $\pm$ 0.01	\\
GS 1124-683 &  7.24 $\pm$ 0.70  & 54.0 $\pm$ 1.5 &  5.89 $\pm$ 0.26 &  -0.24$^{+0.05}_{-0.64}$ & K &  	\\
GX 339-4  &  &  &  &  &  	      & $>$0.97 $^{b}$ \\
GRS 1915+105 & 14.0 $\pm$ 4.4  & 66 $\pm$ 2 & 11.0 & $\ge$0.95 & K2  & 0.98 $\pm$ 0.01	\\
 & & & & $\sim$0.7 $^{c}$ & B & \\
GRS 1739-278 &    &   & &  & & 0.8 $\pm$ 0.2 $^{d}$\\
SAX J1711.6-3608 &   & &  &  &  	      & 0.6 $^{+0.2}_{-0.4}$	\\
Swift J1753.5-0127 &   &  & & &  	      & 0.76 $^{+011}_{-0.15}$	\\
Swift J1910.2-0546 &   &  &  & &  	      & $\le$ -0.32	\\
LMC X-1 & 10.91 $\pm$ 1.54  & 36.4 $\pm$ 2.0 &  48.1 $\pm$ 2.2 & 0.92$^{+0.05}_{-0.07}$ &  K2   & 0.97$^{+0.02}_{-0.13}$ \\
LMC X-3 & 6.95 $\pm$ 0.33 & 69.6 $\pm$ 0.6 &  48.1 $\pm$ 2.2 & 0.25$^{+0.20}_{-0.29}$ $^{e}$ &  	K2      & 	\\
M31 ULX-2 & $\sim$10 & $<$ 60 & 772 $\pm$ 44 & $<$ -0.17 $^{f}$ & B & \\
M33 X-7 & 15.65 $\pm$ 1.45 &74.6 $\pm$ 1.0  & 840 $\pm$ 20 & 0.84$\pm$0.05 & K & \\
\hline
\hline
\end{tabular}
Notes: Spin values from continuum fitting (and system parameters used for the model fits) and from reflection fitting (see Miller \& Miller 2015 for individual references) with errors typically quoted at 1-2$\sigma$ (with the exception of the upper limit for the spin of M31 ULX-2 which is at 3$\sigma$). The differing levels of quoted precision for the values is a result of the individual studies. The model used in the continuum fitting is either {\sc kerrbb: K}, {\sc kerrbb2: K2} or {\sc bhspec: B}. Updated or additional values are indicated by: $^{a}$: Parker et al. (2015), $^{b}$: Ludlam, Miller \& Cackett (2015), $^{c}$: Middleton et al. (2006, as GRS 1915+105 is extreme and spectral modelling is inherently degenerate, we include this value for completeness: see also McClintock et al. 2006 for additional discussion), $^{d}$: Miller et al. (2014), $^{e}$: Steiner et al. (2014), $^{f}$: Middleton et al. (2014). 
\end{minipage} 

\end{center}
\end{table*}





\subsubsection {Reflection fitting}

As mentioned in the previous section, due to the proximity of Galactic BHBs, their X-ray flux is typically very high; unsurprisingly these provided the very first discovery of a broad Fe K$_{\alpha}$ line through an {\it EXOSAT} observation of Cygnus X-1 (Barr, White \& Page 1985; Fabian et al. 1989). As the disc emission does not need to be isolated, even higher Eddington fraction observations than those typically selected for the continuum fitting method may be utilised (although should advection occur at such higher rates, then particle orbits may be affected, e.g. Narayan \& Yi 1994). Due to the source brightness, CCD detector issues may one again become important, but where these distorting effects can be reliably ignored or corrected (see Miller et al. 2010 for a discussion of pile-up on Fe K$_{\alpha}$ line measurements), CCD spectroscopy of BHBs offers the opportunity to study the reflection spectrum in remarkable detail and thereby well constrain the spin.

As discussed in section 3.2, modelling of the reflection spectrum is less sensitive to uncertainties in system parameters such as the distance and BH mass (whilst the inclination is a free parameter in the models) and as such the  method is in principle more robust than modelling the disc emission. However, an important consideration when modelling the reflection spectrum in BHBs is the selection of the continuum which, in the X-ray band of most detectors contains a significant contribution from the disc due to the BH mass (see equation 27). In addition, the nature of the Comptonised emission remains a point of debate, with suggestions from broad-band spectroscopy extending to high energies, that it may be a combination of more than one component due to a two-temperature plasma or a combination of thermal and non-thermal electron populations (e.g. Parker et al. 2015). When considering the spin value obtained via this technique, one must therefore take into account errors resulting from the combination of detector effects and model uncertainty (e.g. the radial profile discussed in Reynolds \& Fabian 2008), and also any uncertainty in the continuum onto which the reflection spectrum is imprinted. Naturally, as the count rate is typically extremely high, these will dominate over over statistical errors. As a final point, it is possible that the Fe K$_{\alpha}$ line may be broadened due to scattering in the disc atmosphere which can confuse measurements of the spin, however as Steiner et al. (2011) find in the case of XTE J1550-564, the effect whilst noticeable, is unlikely to dominate over Doppler and GR effects.

At the time of writing, 17 BHBs (out of a total Galactic population of $\sim$30: Grimm, Gilfanov \& Sunyaev 2002) have measurements for the spin from reflection fitting and these are shown in Table 1 (alongside those from continuum fitting). As discussed in detail by Reynolds (2014), there is general concordance with notable exceptions being 4U 1543-475 and GRO J1655-40. 

The distribution of spin values for stellar mass BHs provides insight into the natal supernova process by which they are formed (see King \& Kolb 1999). As remarked upon by Miller \& Miller (2014), the distribution of values from the two methods are similar (although the sample sizes do not yet allow for Gaussian distributed statistics) with spin values at least two orders of magnitude higher than in the case of neutron stars (where the spin can be accurately determined from pulse periods) and demands vastly different means of acquiring angular momentum during formation.

\subsection{Results: AGN}

Whilst the spin distribution of BHBs is considered relevant for understanding their formation process (King \& Kolb 1999), so too is it the case that the spin distribution for AGN provides information as to the growth of SMBHs (e.g. Moderski \& Sikora 1996; Madau \& Quataert 2004; Volonteri et al. 2005; King, Pringle \& Hofmann 2008; Fanidakis et al. 2011). If the growth progressed via accretion of gas, for example driven by minor mergers with satellite galaxies, then the infalling material will have a range of angular momenta relative to the BH spin (see e.g. Nayakshin, Power \& King 2012) giving a range of SMBH spins in the local Universe. Conversely, accretion of material through the galactic disc with a fixed direction of angular momentum is likely to produce high spins (even if the BH is initially mis-aligned: King et al. 2008). As accretion is inherently connected to the production of highly energetic outflows in the form of winds and jets, the strength of which are potentially related to the BH spin (via the Blandford-Znajek mechanism - see section 4 - and the radiative efficiency in the disc - see equation 30), understanding how SMBHs grew and how their spin evolved is of broad importance for our understanding of the larger scale structure of the Universe due to the interactions of outflows with the host galaxies (see the review of feedback by Fabian 2012). 

In the following sections we discuss the campaigns to determine the spin of AGN excluding those accreting at very low (quiescent levels) such a Sgr A*; although we note that spin estimates via GRMHD simulations and SED fitting for this important SMBH have generally favoured moderate to high (but not maximal) spin values (e.g. Mo{\'s}cibrodzka et al. 2009; Dibi et al. 2012; Drappeau et al. 2013) and will be further constrained when the event horizon telescope is fully operational.


\subsubsection {Continuum fitting}

From inspection of equation 27, it is readily apparent that the peak disc temperature scales inversely with the BH mass. For typical SMBH masses of $>10^{6}$ M$_{\odot}$ (e.g. Woo \& Urry 2002), at sub-Eddington rates, the disc will peak in the extreme UV (Malkan 1983) which is heavily absorbed by the ISM. Whilst substantial emission emerges at higher energies, providing the illuminating continuum for reflection, direct fitting to the continuum to obtain the spin has not been viewed as a promising technique. In addition, for accurate spin measurements to be obtained from the continuum it is important that the system parameters (i.e. inclination, mass and distance: see section 3.1) are known to a high degree of precision. This can be relatively challenging; whilst the distance is known to far greater accuracy than is usually available for Galactic sources (and motivates the study of extragalactic BHBs - Middleton et al. 2014), the mass and inclination are harder to measure with typical errors on the mass via reverberation estimates of 0.5 dex. 

A subclass of AGN - specifically some narrow line Seyfert 1s (NLS1s) - appear to show a very hot disc component (e.g. Middleton et al. 2007; Jin et al. 2012a, 2012b; Terashima et al. 2012) which may extend into the soft X-ray band and contribute to the ubiquitous `soft excess' (Czerny et al. 2003; Gierli{\'n}ski  \& Done 2004) directly or via Comptonisation (Done et al. 2012). Recent progress in modelling the AGN inflow - most notably by applying a more rigourous treatment for the opacity balance (see section 3.1) - has lead to the development of a model which incorporates approximate radiative transfer, {\sc optagnf}, somewhat analogous to {\sc bhspec} (Done et al. 2012). This model takes in the mass accretion rate which can be determined from the optical flux where the mass is reasonably well estimated (Davis \& Laor 2011) and conserves energy extracted from the accretion process in creating an optically thick thermal Compton component and the higher energy tail. Done et al. (2013) applied this new model with a relativistic convolution kernel ({\sc kerrconv}) to the spectrum of the bright NLS1, PG 1244+026. The spectrum of this source is similar to other extreme, soft NLS1s (e.g. RE J1034+396: Middleton et al. 2009), appearing to have high (close to the Eddington limit) mass accretion rates with a very weak tail of hard emission and a strong soft-excess. Whilst high Eddington accretion rates are not generally considered to be an appropriate regime to apply standard continuum fitting for BHBs (see section 3.1), analogous behaviours of the AGN disc have not been established (due to the timescales of variability). As this model conserves the energy produced via accretion (and is therefore related to the spin) in creating the regions of different optical depth, the limits on the spin imposed by this model are therefore of interest (although as with more standard models does not yet take into account energy lost in the creation of a wind or jet). In the case of PG 1244+026, Done et al (2013) find that the spin must be low (and rule out maximal spin). Such an approach is likely to be useful for those lower mass AGN where the broad-band (optical to X-ray) disc emission can be well modelled, i.e. relatively unobscured by gas and dust. Whilst these appear relatively rare (see Middleton et al. 2007), the future missions of {\it eROSITA} and {\it Athena} are expected to find many more, allowing this approach to be more widely used and thoroughly tested.

\subsubsection {Reflection fitting}

The vast majority of spin measurements for AGN, have come from fitting reflection models to their spectra. By AGN in this context we are generally referring to Seyfert 1 AGN (including NLS1s and QSOs) as the unified model (Antonucci \& Miller 1985; Antonucci 1993) would suggest that these are viewed at low to moderate inclinations and so are not obscured by the molecular torus (which is the origin of Compton-thick AGN where Fe K$_{\alpha}$ emission is seen but originates from the torus and so is not a measure of the BH spin). The review of Reynolds (2014) provides details of the approaches discussed already as well as a `cookbook' for obtaining estimates for the BH spin from the reflection spectrum and we direct the interested reader and practical observer here.  

The Fe K$_{\alpha}$ line is known to be almost ubiquitous in AGN (Nandra et al. 1997; Reynolds 1997) as a direct result of X-ray illumination of optically thick (not fully ionised) material and was first discovered  by Tanaka et al. (1995) when the bright AGN, MCG-6-30-15, was observed by ASCA with the strong, broad line found to require maximal spin (see also Iwasawa et al. 1996; Dabrowski et al. 1997; Reynolds \& Begelman 1997; Young, Ross \& Fabian 1998). Since the inception of the field, the use of the reflection spectrum (the whole of which is pivotal for accurate spin determination: Young, Ross \& Fabian 1998) has been widespread, finding an application in probes of higher redshift, lensed QSOs (Reis et al. 2014; Walton et al. 2015) as well as in novel techniques to utilise the time dependence of the emission (Risaliti et al. 2011 - see section 5).

There are a number of important considerations when fitting the reflection spectrum of AGN which are not relevant in the application of this method to BHBs. The optical/UV line ratios of Seyfert AGN (e.g. Warner, Hamann \& Dietrich 2004; Nagao, Maiolino \& Marconi 2006) imply that the metal abundance of the gas is super-solar (e.g. Zogbhi et al. 2010; Reynolds et al. 2012) which plays an important role in the strength of the Fe K$_{\alpha}$ line; it is therefore important to allow any applied models to extend beyond solar metallicities. Unlike accretion discs around stellar mass BHs, the discs around SMBHs do not contribute in a sizeable way to the X-ray continuum (with the exception of the very brightest, lowest mass AGN: Middleton et al. 2007; Jin, Ward \& Done 2012; Done et al. 2012) giving a much cleaner view of the reflection spectrum. However, the environment of the AGN is less ``clean" than in BHBs: winds are expected to be ubiquitous given the UV radiation field and resonant line opacity (Proga \& Kallman 2004), and will not be as ionised as those from BHBs. These winds can therefore result in a distorting imprint around the Fe K$_{\alpha}$ line, and possible degeneracy in the spectral fitting of AGN (e.g. Patrick et al. 2012; Brenneman et al. 2012). This degeneracy can be broken in two separate ways, the first is through use of an observable bandpass which extends to higher energies, where differences between absorption and reflection differ dramatically (as demonstrated in the case of NGC 1365 through use of {\it NuSTAR} data: Risiliti et al. 2013). The second means to break the spectral degeneracy is through use of variability techniques (e.g. the cross spectrum: Nowak \& Vaughan 1997) which can effectively isolate reverberation of the primary continuum (Uttley et al. 2014). The latter technique has clearly demonstrated that the lags at high frequencies (i.e. those originating from the most compact regions of the corona) contain a signature of the broad Fe K$_{\alpha}$ line (Zoghbi et al. 2012; Kara et al. 2013) and Compton hump (Kara et al. 2015a; 2015b), confirming the strong contribution of reflection to the spectrum.

A further important consideration when applying the reflection fitting method to AGN is the presence of the soft excess. As the name implies, this is an excess of flux seen below 2~keV once a power-law fit to the 2-10~keV band is extrapolated backwards and has no obvious counterpart in BHB spectra. The soft excess is both smooth and in present across a large number of AGN across a range in mass, peaking at $\sim$0.5~keV in each; this raises problems for its interpretation as a continuum-only component should not produce the same peak temperature across a range of masses without some fine-tuning of the radiative process (Gierli{\'n}ski  \& Done 2004). Instead atomic transitions associated with OVII/OVIII and the Fe M UTA naturally produce features in this energy range, and could be seen in absorption and emission via reflection or in an outflow. However, in order to produce such an excess of flux without seeing sharp features requires velocity broadening in either situation. Crummy et al. (2006) and Middleton et al. (2007) studied a sample of AGN, finding that both reflection and absorption can produce statistically indistinguishable fits (across the {\it XMM-Newton} bandpass) with the spin tending towards large values and the outflow velocity of the wind tending to being moderately relativistic. A third possibility is the combination of reflection, outflow and, in the case of the brightest, low mass AGN, some Compton component of the disc (Done et al. 2012; Jin et al. 2013). Understanding the origin of the soft excess is considered to be extremely important as, in cases where the inclination of the disc is unknown or cannot be constrained in the reflection model, the spin can be driven by the need to smear the atomic features at soft energies. Once again, applying methods which utilise the time domain can assist in understanding the origin of this component and have shown that for some sources the likely origin is in partially ionised reflection (notably 1H0707: Fabian et al. 2009; Zoghbi et al. 2010) whilst in others it would appear that the soft excess is dominated by a continuum component associated with Compton up-scattering of UV seed photons (Jin et al. 2013).

In the case where the inclination is well constrained by the model (and therefore the result does not rely on the soft excess) and the whole reflection spectrum is considered (allowing for super-solar abundances of metals), the uncertainty on the spin value once again depends on the continuum (including the effect of absorption) and the model  being used, whilst at typical AGN fluxes, detector issues which can be challenging for observations of BHBs become less troublesome.


In general, SMBH spins determined via reflection (see for example the sample studies of Walton et al. 2013 and Patrick et al. 2012) show a tendency towards high spin (as remarked upon by Reynolds 2014). As discussed by Brenneman et al. (2011) and Walton et al. (2013), this distribution may be a result of selection effects; a higher spin results in higher radiative efficiency (see equation 30) and a brighter AGN; conversely this may be indicating coherent rather than chaotic SMBH growth which would result in systematically high spins (see Fanidakis et al. 2011 although see Nayakshin, Power \& King 2012 for counter arguments).

\subsection{Implications: Powering of ballistic jets}

As remarked upon in section 2.1 of this chapter, the effect of frame-dragging will have important consequences for the transfer of energy from the BH to an orbiting test particle. Here we briefly describe the Penrose process (Penrose 1969; Christodolou 1970; Bardeen et al. 1972) and how it can be linked via the Blandford-Znajek mechanism to the powering of relativistic ejections.

\subsubsection{The Penrose effect \& Blandford-Znajek mechanism}

The mechanical Penrose effect occurs as a result of `negative energy orbits', i.e. where the energy required to send a body to infinity is larger than its rest mass energy. In the notation of GR this is u$_{t} <$ 0 and can be shown to only be possible for u$_{\phi} >$ 0 and:

\begin{equation} 
(u_{\phi})^{2}(-g_{tt}) > \Delta sin^{2}\theta
\end{equation}
where $\Delta$ is given in equation 4. As the right-hand side of the inequality is positive, we find that this is only satisfied if g$_{tt} <$ 0, i.e. within the ergosphere. It can then be shown that for a particle to have negative energy, i.e. u$_{\phi} >$ 0, orbits need to be retrograde relative to the BH spin. The Penrose process can then be described as a body entering the ergosphere and breaking apart; one part is induced into a retrograde orbit and the other escapes to infinity after having gained energy (equal to the negative energy captured by the BH) from the rotation of the BH. The efficiency of this process, i.e. the ratio of maximum energy out to that going in, is of the order 20\% (Chandrasekhar 1983, see also the review of Brito, Cardoso \& Pani 2015).

Whilst the mechanism described above is unlikely to have a direct impact on the observational appearance of BHs, when coupled to magnetic fields, it may present a viable mechanism for powering collimated, relativistic outflows in the form of jets. Poloidal magnetic fields are expected to grow in the accretion flow via the magneto-rotational instability (MRI: Balbus \& Hawley 1991) and propagate down to the vicinity of the BH. Where these field-lines thread the ergosphere they are forced to rotate with the matter, inducing a force on the coupled charged plasma (Lorentz force) which will lead to acceleration of material at relativistic speeds along the rotation axis of the BH in the form of jets (which are then collimated via magnetic confinement). This is a highly simplified description of the Blandford-Znajek (BZ) mechanism (or effect: Blandford \& Znajek 1977) where the power that can be extracted P$_{\rm jet} \propto a^{2}$ as long as the spin is not large. A more accurate relationship that covers the whole range of possible spins has been derived by Tchekhovskoy \& McKinney (2012) to be P$_{\rm jet} \propto (M\Omega_{\rm H})^{2}$ where $\Omega_{\rm H}$ is the BH angular frequency which (in natural units) is given by:

\begin{equation}
\Omega_{\rm H} =  \frac{a}{2M(1 + \sqrt{1 - a^2})}
\end{equation}

Measuring both the spin and jet power in a reliable way can therefore provide insights into the launching of the jets by confirming the BZ-effect (or an effect with a similar form) or by ruling it out.



\subsubsection{Testing for the Blandford-Znajek mechanism}

As briefly mentioned in section 3.3, BHBs launch jets at low bulk Lorentz factors ($\Gamma <$ 2) when accreting at low to high rates and when accompanied by an X-ray spectrum dominated by a hard thermal tail of emission. The emission from the jet extends from low (MHz) frequencies up to optical/IR and is typically a flat power-law (S $\propto$ E$^\gamma$ where $\gamma \approx$  0) across the intervening several decades in frequency. The spectrum results from highly (ultra-relativistic) electrons spiralling around magnetic field lines and cooling via synchrotron radiation. The flat spectrum is a result of viewing the emission from spatially extended regions, with the low frequency radio emission originating further from the launching point (i.e. where the particle Lorentz factor is lower). A break at low frequencies occurs when the lowest energy electron population becomes optically thick to their radiation (and are self-absorbed) whilst a high frequency break occurs when the most energetic electrons are optically thin to their radiation (in reality these breaks occur throughout the spectrum and merely convolve to give the observed spectrum). Although the spectrum evolves (notably the position of the high frequency break as the electrons cool, e.g. Russell et al. 2015), as the emission is constant, the jet is termed `steady'.

The jet changes dramatically in nature when the spectrum softens into the intermediate and then soft states, gaining a much higher bulk Lorentz factor ($\Gamma > $2) and taking the form of discrete ejecta (e.g. Mirabel \& Rodriguez 1999,  Fender et al. 2009) which cool via synchrotron radiation with a high frequency spectral break evolving with the expansion of the ejecta (see van der Laan 1966; Kellermann \& Owen 1988; Hjellming 1988; Hjellming \& Johnston 1988). These ejections were first observed in the Galactic centre `Annihilator', 1E 1740.7-2942 (Mirabel et al. 1992) and subsequently detected in the remarkable BHB, GRS 1915+105 which was the first Galactic source where superluminal jets (appearing as such due to their highly relativistic velocities and orientation close to the line-of-sight) were identified; due to their resemblance to the superluminal ejections from radio loud Quasars, these sources were dubbed `microquasars' (Mirabel \& Rodriguez 1994). Such discrete ejections are thought to be ubiquitous, with {\it all} BHBs showing (or expected to show) what are sometimes referred to as ``ballistic" jets.

Typically the power in the steady jet is determined from the established correlation with the radio synchrotron luminosity (Blandford \& K{\"o}nigl 1979; K{\"o}rding, Fender \& Migliari 2006; Jamil, Fender \& Kaiser 2010), whilst the power in the ballistic jet has traditionally been estimated from the monochromatic radio luminosity which Steiner et al. (2013) have shown to be approximately linearly correlated with the mechanical (bulk kinetic) jet power (see their Appendix). To obtain the {\it intrinsic} luminosity, the effect of Doppler boosting (which acts on both the flux and energy of any breaks in the synchrotron spectrum) has to be accounted for. In practice the boosting factor (D$^{3-\gamma}$: see equation 34) is determined from the inclination (although, as discussed in section 2.1.1, one has to careful as to which inclination - inner or outer disc - is being used) for a range of bulk Lorentz factors and assuming a typical spectral index. By correlating this de-boosted jet power against the spin derived from the spectrum using the continuum fitting or reflection fitting methods, the likely impact of the BZ (or a similar) mechanism can then be tested. 

Fender, Gallo \& Russell (2010) demonstrate that there is no obvious correlation between the steady jet power and the spin for a sample of BHBs, effectively ruling out the BZ effect as the dominant mechanism for the launching of the slower jet. To investigate the powering of the faster, ballistic jet, Narayan \& McClintock (2012) and Steiner, McClintock \& Narayan (2013) selected the five Galactic BHBs which are thought to reach their Eddington limit and therefore act as `standard candles', thereby removing any mass accretion rate dependent effects; the resulting presence of a correlation between jet power and spin has been claimed as strong support of the BZ effect. Russell, Gallo \& Fender (2013) have since questioned the selection criteria arguing that the inclusion of other sources disagrees with the presence of a correlation; instead the driving factor is claimed to be the mass accretion rate (with similar arguments proposed for AGN: King et al. 2013; 2015). Whilst the debate is ongoing, it is abundantly clear that a larger sample of BHBs is required in order to fully evaluate the presence of a correlation. The application of these techniques to nearby extragalactic BHBs (which can be studied reliably in the X-ray and radio: Middleton et al. 2013) is in its infancy but is the only means by which this can be accomplished (Middleton et al. 2014). In addition, the introduction of new methods for measuring the spin (as discussed in the following sections), will provide important cross-checks for existing methods and provide increased confidence in any conclusions which rest upon its measurement. 

As a final remark, Garofalo et al. (2014) note that the BH spin itself may have a distorting influence on the nature of any correlation between jet power and spin through the quenching of the jet by winds which become stronger with increasing source brightness (e.g. Neilson \& Lee 2009). This assumes that the winds are radiatively driven (either thermal via reprocessing or radiation pressure powered via scattering) such that a higher spin - which leads to a higher efficiency in conversion of rest mass to energy via equation 30 - more readily powers winds and could in principle shut off the jet at lower mass accretion rates (under the assumption that the wind launching is not dependent on the mass accretion rate dependent structure of the disc itself). Whilst the exact mechanism for jet quenching is not yet fully understood, certainly the coupled interaction of inflow and outflow and the spin dependence is an important consideration for understanding the evolution of accreting BHs of all masses (e.g. Kov{\'a}cs, Gergely \& Biermann 2011).

\subsection {Implications: Retrograde spins?}

In a very small number of cases: only 3 BHBs (Reis et al. 2013; Middleton et al. 2014; Morningstar et al. 2014) 
to date, significantly retrograde ($a_{*} < 0$) spin has been reported. It is possible that such sources could launch particularly powerful jets should magnetic flux be swept from the plunging region onto the BH, with the amount of flux being dependent on the size of the `gap' between the ISCO and BH. As this is larger for retrograde spin (R$_{ISCO}$ = 9R$_{g}$ for a$_{*}$ = -0.998, see Figure 2), the magnetic flux trapped on the BH can therefore be enhanced (Garofalo 2009; Garofalo, Evans \& Sambruna 2010) although simulations incorporating the effect of magnetic field saturation (Tchekhovskoy \& McKinney 2012) dispute this `gap paradigm model' and arrive at the opposite conclusion. 

Irrespective of the impact, we must ask the question, how can retrograde spin practically occur in those systems where it has been reported and how likely is it? Co-alignment of disc and BH (which occurs through the action of viscous torques transferred via the Lense-Thirring and Bardeen-Peterson effects, e.g. King et al. 2005) is expected to take at least several percent of the binary's lifetime and so, assuming that the measurement of retrograde spin is genuine (and not an artefact of inner disc truncation due to some as yet unknown process), retrograde spin is likely to be either an indication of the formation process (i.e. an anisotropic supernova kick: Brant \& Podsiadlowski 1995), the result of wind-fed accretion, (which can produce counter-aligned inflows, e.g. GX 1+4: Chakrabarty et al. 1997 and Cyg X-1: Shapiro \& Lightman 1976, Zhang et al. 1997) or from the tidal capture of a star (Fabian, Pringle \& Rees 1975). In this last case, after the BH has formed whilst in a globular cluster, it is expelled by the natal kick, with a subsequent stellar capture producing a retrograde orbit. Whether retrograde spin could form in AGN is not clear but could presumably result from minor mergers with material carrying counter-aligned angular momentum accreted onto the SMBH on short time-scales (as on long timescales enough material can be accreted to co-align the system). Notably in AGN, the entire sub-pc disc may be misaligned with the Galaxy as a result of a recent minor merger and indeed there is growing evidence for this from jet launching angles and the location of the molecular torus (see Hopkins et al. 2012).


\section{Observational tests of spin II - the time domain and relativistic precession model}

Traditional methods of determining the spin have proven to be highly illuminating, however methods which do not rely solely on the time-averaged spectrum have the advantage of being able to provide a semi-independent measure of the spin and test traditional models (and our understanding of GR). 

As we discussed in section 2.1.1, the effect of relativistic frame-dragging leads to precession of orbits in the innermost regions of the accretion flow due to the Lense-Thirring effect. Such precession in turn leads to epicyclic oscillations about the vertical `nodes' (where orbits of a test particle out of the equatorial plane meets that of the ecliptic) and precession of periastron passage. When combined with the frequency of Keplerian orbit, these three frequencies form the relativistic precession model (or RPM: Stella \& Vietri 1998; Stella, Vietri \& Morsink 1999; Stella \& Vietri 1999). As long as the nature of the plasma in the accretion disc is not so dense as to dampen the oscillations, in practice we should expect each of these frequencies to leave a trace of coherent power in the lightcurve of accreting back hole systems (both BHBs and AGN). These could naturally be associated with the quasi-periodic oscillations (QPOs) detected as narrow peaks in the power density spectrum (PDS) and seen in BHBs at low frequencies (e.g. Wijnands, Homan \& van der Klis 1999; Casella, Belloni \& Stella 2005; Belloni, Motta \& Mu{\~n}oz-Darias 2011; Motta et al. 2012), occasionally (and only detected so far in 5 BHBs) at high frequencies (e.g. Morgan, Remillard \& Greiner 1997; Remillard \& McClintock 2006; M{\'e}ndez  et al. 2013) and recently discovered in AGN (Gierli{\'n}ski  et al. 2008; Middleton et al. 2009, Middleton, Uttley \& Done 2011; Alston et al. 2014; 2015). The AGN QPOs appear to be analogous to the high frequency QPOs (HFQPOs) of BHBs (Middleton \& Done 2010) and, in both sets of systems, appear in a 3:2 harmonic ratio implying a common physical origin (e.g. Dexter \& Blaes 2014).


In the RPM, the two HFQPO frequencies are associated with the orbital frequency, $\nu_{\phi}$,
and the periastron precession frequency, $\nu_{per}$, which in turn is given by the difference between the orbital and radial epicyclic frequencies: $\nu_{per} = \nu_{\phi} - \nu_{r}$. The  much slower, vertical (Lense-Thirring) precession would instead be associated with the low frequency QPO (LFQPO: Ingram et al. 2011; Ingram \& Done 2012) and is given by $\nu_{lt} = \nu_{\phi} - \nu_{\theta}$  (where $\nu_{\theta}$ is the vertical epicyclic frequency). The three QPO frequencies are connected both to one another and to the BH mass and spin through the following relations (Bardeen, Press \& Teukolsky 1972; Merloni et al. 1999):

\begin{equation}
\nu_{\phi} = \pm\frac{\beta}{M}\frac{1}{r^{3/2}\pm a}
\end{equation}

\begin{equation}
\nu_{per} = \nu_{\phi}\left[1 - \sqrt{1 - \frac{6}{r}\pm\frac{8a}{r^{3/2}} - \frac{3a^{2}}{r^{2}}}\right]
\end{equation}

\begin{equation}
\nu_{lt} = \nu_{\phi}\left[1 - \sqrt{1 \mp \frac{4a}{r^{3/2}} + \frac{3a^{2}}{r^{2}}} \right]
\end{equation}
where $\beta = c^{3}/(2\pi GM_{\odot})$, r is the radius in units of gravitational radii and M is in units of solar mass. Where $\pm$ or $\mp$ are given, the top sign refers to a treatment where the spin is prograde and the bottom sign to where the spin is retrograde.

The above set of equations lead to a set of simultaneous equations and in turn to the following formula for the spin (see Ingram \& Motta 2014 for a derivation):

\begin{equation}
a = \pm \frac{r^{3/2}}{4}\left[\Lambda + \Phi - 2 + \frac{6}{r}\right]
\end{equation}
where:

\begin{equation}
\Phi = \left(1 - \frac{\nu_{per}}{\nu_{\phi}}\right)^{2}
\end{equation}
and

\begin{equation}
\Lambda= \left(1 - \frac{\nu_{per}}{\nu_{lt}}\right)^{2}
\end{equation}

Although the solutions to the equations of the RPM do not immediately tell us whether the spin is prograde or retrograde, Ingram \& Motta (2014) point out that this can be identified from the highest frequency reached by the LFQPO (which if it extends to within the ISCO for $a = -|a|$ then implies prograde spin).

Critically, the application of the RPM in determining the spin relies on the {\it simultaneous} presence of QPOs in the lightcurve (as their frequencies are indicative of the radius at which they are generated), although it is not vital that all three be present as Ingram \& Motta (2014) demonstrate semi-analytically. Recently, the RPM method has been applied to two BHBs: GRO J1655-40 (Motta et al. 2014a), the only BHB to date where all three QPOs have been detected simultaneously (Motta et al. 2012) and XTE J1550-654 (Motta et al. 2014b) where only the LFQPO and one of the HFQPOs have been detected simultaneously. The resulting spin for XTE J1550-654 was found to be consistent with estimates from the reflection fitting and continuum method (Steiner et al. 2011) whereas the value obtained for GRO J1655-40 via the RPM is inconsistent with those obtained via spectral means (Shafee et al. 2006; Miller et al. 2009; Reis et al. 2009). The reason for the discrepancy in the spin for the latter source may be due to the misalignment of the BH spin and orbital axis by $>$ 15$^{\circ}$ (Greene et al. 2001) which are likely to be closer aligned in XTE J1550-654 (Steiner \& McClintock 2012). Thus, whilst the RPM is insensitive to the inclination (as the frequencies are independent of the inclination unlike the measured strengths of the QPO: Motta et al. 2015), this could present problems for measurements of the spin via spectroscopic methods (see the discussions in section 3).

Finally, it should be pointed out that mechanisms to explain the origins of the QPOs besides the RPM have also been proposed (e.g. Esin, McClintock \& Narayan 1997; Taggar \& Pellat 1999), with little {\it direct} observational progress made in distinguishing the correct interpretation. However, a key discriminator for the origin of the LFQPO as Lense-Thirring precession may be the QPO phase-resolved reflected emission (see section 3.2) as described in Ingram \& Done (2012), where the geometrically thin, optically thick disc truncates (possibly due to disc evaporation, e.g. Meyer, Meyer-Hofmeister 1994; Liu et al. 1999) and the inner, lower density region (in this picture, the location of the Compton up-scattering of seed disc photons) precesses as a solid body as a result of the Lense-Thirring effect (see Ingram, Done \& Fragile 2009). As the inner region precesses, various disc azimuths are subjected to changing illumination; in turn this leads to changes in the observed reflected emission (Fe lines and Compton hump) as a result of the various Doppler-shifts and boosting. Should these predicted changes be observed by long observations with present instruments or using high throughput, high time-resolution instruments such as the LAXPC onboard the recently launched {\it ASTROSAT} (see Ingram \& van der Klis (2015) for details of the arithmetic approaches and a possible detection of modulation already seen in {\it RXTE} data), it will likely represent a `smoking gun' for the RPM (and may likewise rule out such an origin if the predicted variations are not observed).


\section{Observational tests of spin III - the energy-time domain}

As opposed to the previously described methods which use either the energy {\it or} time domain, an approach which combines the two promises to provide the largest lever arm for estimating the BH spin. One such method is referred to as `Doppler Tomography' and relies upon a changing view of the regions of the accretion disc due to an eclipse by an orbiting body. This technique was first applied to the study of white dwarfs in mapping the accretion disc via emission lines (e.g.  Marsh \& Horne 1988) and a small number of authors have since developed it as a tool for the study of AGN spin and the effects of GR.  

In the majority of AGN (with masses typically $>$ 10$^{6}$ M$_{\odot}$: Woo \& Urry 2002) the disc is out of the X-ray bandpass (although the hottest tail of the disc and/or a Compton up-scattered component may enter at soft energies for the lowest mass and highest mass accretion rate sources: Middleton et al. 2007; Jin et al. 2011; Done et al. 2012, see the discussion in section 3.1) whilst the primary Compton scattered emission and its reflected component dominates the emission (e.g. Fabian 2009). Should an orbiting body pass across our line-of-sight, an eclipse results and leads to changes in the spectrum as a function of time that allows a test of the nature of the inner regions, e.g. the radial temperature dependence of the corona, and spin (see McKernan \& Yaqoob 1998).

Variability due to obscuration by cold material is relatively common on long timescales in AGN (see Risaliti, Elvis \& Nicastro 2002) and on shorter timescales by Compton-thin material (Elvis et al. 2004; Puccetti et al. 2007; Bianchi et al. 2009; Risaliti et al. 2010, 2011a). However, obscuration by Compton-thick (i.e. $\tau >$ 1 $\rightarrow n_{H}\sigma_{T} > $1 $ \rightarrow n_{H} > $ 10$^{24}$ cm$^{-2}$) material on observable ($<$ 100s of ks) timescales - which leads to the simplest form of eclipse - is relatively rare, although at least one such event has been observed in NGC 1365 (Risaliti et al. 2009).  


\begin{figure}[h]
\sidecaption
\hspace{1.5cm}
\includegraphics[scale=.6]{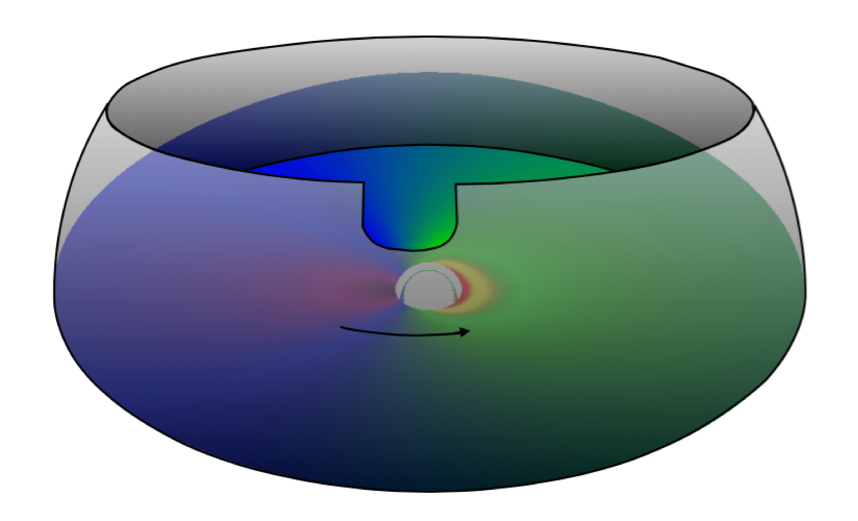}
\caption{Schematic of a wind showing a gap (assumed to have been formed via Rayleigh-Taylor or similar instabilities), rotating and providing a changing view through to the inner disc regions (the emission from which is determined from ray-tracing: Dexter \& Agol 2009). Analogous to the situation of eclipses by Compton-thick clouds (Risaliti et al. 2011b), the emission seen by the observer as a function of time is dependent on the inclination and spin (as well as disc structure). By incorporating the additional lever arm of timing analysis,  Doppler Tomography provides a means to obtain tighter constraints on the spin than is possible through use of the time-averaged spectrum alone (Middleton \& Ingram 2015).}
\label{fig:2}       
\end{figure}

Using the model of Dov{\v c}iak et al. (2004), which calculates the line emission from different parts of the disc separately, Risaliti et al. (2011b) simulate the effects of a eclipse under the assumption that the obscuring source is a Compton-thick broad-line region cloud and completely covers the source. The result is a profound shift in the shape of the Fe K$_{\alpha}$ line as a function of time as the approaching blue-shifted side of the disc is covered followed by the retreating red-shifted side (i.e. a situation where the cloud is co-rotating with the disc). As Risaliti et al. (2011b) point out, the major effect is not a shift in the line profile but in flux due to Doppler boosting. As a corollary, the high-energy continuum emission, i.e. the Compton hump, will also rise and fall in flux, correlated with changes in the emission line (Figure1 of Risaliti et al. 2011b). Risaliti et al. (2011b) point out that this would constitute a case of a `perfect' eclipse as the obscuration is complete and the eclipse assumed to have sharp, linear edges. The authors perform simulations which show that present observatories have the required throughput to detect predicted changes for Compton-thick eclipses and can provide independent confirmation of relativistic effects in shaping the Fe K$_{\alpha}$ line. Although the more common eclipses by Compton-thin material (typically a few 10$^{23}$ cm$^{-2}$: Risaliti et al. 2009; Maiolino et al. 2010) have less of an impact on the reflection spectrum, Risaliti et al. (2011b) show that future observatories (for instance ESA's {\it Athena}) will still be able to detect changes associated with the passage of material. 

Although not modelled explicitly, the change in Doppler boosting from either side of the disc is dependent on both the inclination to the source and rotational velocity (see equation 34). The latter is of course dependent on the location of the ISCO and therefore the spin. Such a technique therefore not only provide a means to independently verify the origin of the emission in relativistic material but also measure the spin. There are of course qualifiers and caveats to this approach, and Risaliti et al. (2011b) point out that variations in the intrinsic emission can lead to a distorting effect as can an eclipse that covers the illuminating source in a different manner to the reflector. 

An analogous situation to that described by Risaliti et al. (2011b) can be applied to the disc emission directly when eclipses take place in AGN of lower mass. Such low mass AGN are preferentially detected at the highest mass accretion rates (Greene \& Ho 2004), where, analogous to the spectra of BHBs (Jin et al. 2012a; 2012b), the spectrum is dominated by the disc with a weak, flat power-law tail of emission to high energies. In such sources, reflection features are therefore expected to be weak, restricting the means by which the spin can be measured. However, at the apparent high accretion rates associated with low-mass AGN, powerful winds are expected to be driven from the disc, which itself may have grown in scale-height (e.g. Shakura \& Sunyaev 1973; King 2001; Poutanen et al. 2007). As such winds are likely to rotate in an approximately Keplerian manner (with deviations from this expected to scale as (H/R)$^{2}$), any inhomogeneities due to radiative-hydrodynamic instabilities in the surface of the wind material (Proga \& Kallman 2004) will lead to gaps through to the inner regions which also rotate. Should our sight-line to the source intercept one of these gaps, we can obtain a view of the approaching, blue-shifted side of the disc and then the retreating red-shifted side as the gap orbits (see Figure 5). As with the model presented by Risaliti et al. (2011), this form of Doppler Tomography is highly sensitive to the spin and inclination and can therefore be used to provide independent constraints. Notably, unlike studies of the reflection, the disc emission is expected to be stable over the timescales of an observation and so is not likely to be affected by intrinsic variability that can have a distorting influence.  

Using the ray-tracing code {\sc geokerr} (Dexter \& Agol 2009),  Middleton \& Ingram (2015) create a model to describe the orbit of a gap in a Compton-thick wind and apply it to the case of the low mass AGN RX J1301.9+2747 (Sun et al. 2013) which shows long-lived flaring behaviour (Dewangan et al. 2000) inconsistent with the usual origin of rapid variability in AGN (i.e. viscous - see Uttley, McHardy \& Vaughan 2005 - and/or thermal). The authors instead argue that the variability is due to gaps in the Compton-thick wind crossing our line-of-sight (Figure 5). From fitting the model across multiple phases simultaneously, the authors find the spin to be very low irrespective of several caveats (e.g. errors on the mass and temperature profile in the disc) although once again important assumptions remain including the unknown structure of the disc seen through the gaps (the ray-tracing assumes a Novikov-Thorne disc). Importantly, the combination of the time and energy domains leads to stronger constraints on the spin than can be obtained from traditional methods, demonstrating the power of Doppler Tomography as a method to probe AGN accretion and the region of strong gravity.

%
%

\section{Concluding remarks and future approaches}

In this chapter we have discussed the core theory that is useful for an appreciation of the role BH spin plays, notably the effect of precession and the Penrose process due to frame-dragging and the changing position of the ISCO. As a result of the latter's effect on the emergent radiation (be it direct or reflected), the community has been provided with a means to measure the BH spin in both AGN and BHBs. 

Campaigns over the last 10 years have started to allow the spin distributions of BHs to be probed, allowing progress to be made in understanding their formation. However, many questions remain open and as yet unanswered: What role does the spin play in the launching of ballistic jets? Is there a bias in the spin measurements of AGN or is the spin genuinely high in most local Seyferts? Is retrograde spin common or vanishingly rare? How reliable are our present set of techniques?

The first three questions can only be addressed by expanding our sample of sources for which we have accurate spin measurements. This will no doubt be possible in the forthcoming years when new, highly sensitive X-ray satellites including {\it Athena} and {\it eROSITA} (and potentially {\it LOFT} or a descendent) become available. These will provide the deepest views of the X-ray Universe, providing access to not only the spins of local sources but also, in the case of {\it Athena}, the cosmic evolution of the AGN spin distribution (which in turn probes the growth mechanism of SMBHs: Volunteri et al. 2005; Fanidakis et al. 2011). The photon rich spectra that {\it Athena} will obtain will not only provide high precision spin measurements but potentially even test for deviations from the Kerr metric (e.g. Jiang, Bambi \& Steiner 2015). 

As we have discussed in section 3.5, to rigourously probe the BZ effect (in BHBs - the analogy to AGN jets is still not clear) requires a much larger sample of sources with reliable measure of both BH spin and jet power - for this we must look to nearby galaxies. Such an approach has been proven to be feasible with current instrumentation  (Middleton et al. 2014) and in future will benefit from the introduction of both high throughout X-ray instruments and also the next generation of radio telescopes ({\it SKA} and pathfinders), for which the discovery of radio transients is a core aim (e.g. the ThunderKAT campaign). 

Finally to test the reliability of our techniques requires the use of the time domain in an independent (RPM) or complimentary (Doppler Tomography) fashion and, looking to the future the use of X-ray polarimetry and gravitational wave interferometry. Expanding briefly on the latter two techniques, as explained in Schnittman \& Krolik (2009), the effect of returning radiation from the accretion disc leads to scattering which is not appreciable at low energies (i.e. further out in the disc) and leads to horizontal polarisation (Chandrasekhar 1960), whilst at higher energies the increased scatter to the observer results in vertical polarisation (Agol \& Krolik 2000). The amount of the latter is dependent on the position of the ISCO and therefore the spin. This is expected to be an extremely powerful technique and is the focus of a number of proposed (e.g. {\it PRAXyS} and {\it X-Calibur}) and accepted ({\it XIPE}) missions. The impact of gravitational wave interferometry on the field of BH spin measurements has been discussed in detail in the recent review by Miller \& Miller (2014) to which we point the interested reader; in essence, should a BH-NS or BH-BH binary be found and if the BH spin is high and misaligned with the orbital axis (due to LT precession: see section 2.1), then there can be a considerable impact on the gravitational waveform. 

In conclusion, the future looks extremely bright for the field of BH spin determination and in years to come will allow the most detailed understanding of the most extreme objects in the Universe.

\begin{acknowledgement}
The author gratefully acknowledges the assistance of Chris Reynolds and Jack Steiner in proof-reading and offering valuable suggestions. 
\end{acknowledgement}

\end{document}